\documentclass[sn-mathphys-num]{sn-jnl}

\usepackage{graphicx}%
\usepackage{multirow}%
\usepackage{amsmath,amssymb,amsfonts}%
\usepackage{amsthm}%
\usepackage{mathrsfs}%
\usepackage[title]{appendix}%
\usepackage{xcolor}%
\usepackage{textcomp}%
\usepackage{manyfoot}%
\usepackage{booktabs}%
\usepackage{algorithm}%
\usepackage{algorithmicx}%
\usepackage{algpseudocode}%
\usepackage{listings}%

\usepackage{lmodern}
\usepackage{anyfontsize}

\begin{document}

\title{Effects of Antivaccine Tweets on COVID-19 Vaccinations, Cases, and Deaths}


\author*[1,2]{\fnm{John} \sur{Bollenbacher}}\email{jmbollenbacher@rti.org}

\author*[1]{\fnm{Filippo} \sur{Menczer}}\email{fil@iu.edu}

\author[1]{\fnm{John} \sur{Bryden}}\nomail

\affil[1]{\orgdiv{Observatory on Social Media}, \orgname{Indiana University}, \orgaddress{\street{1015 E 11th St}, \city{Bloomington}, \postcode{47408}, \state{IN}, \country{USA}}}

\affil[2]{\orgdiv{Center for Data Science and AI}, \orgname{RTI International}, \orgaddress{\street{3040 E Cornwallis Rd}, \city{Durham}, \postcode{27709}, \state{NC}, \country{USA}}}


\abstract{
Despite the wide availability of COVID-19 vaccines in the United States and their effectiveness in reducing hospitalizations and mortality during the pandemic, a majority of Americans chose not to be vaccinated during 2021. Recent work shows that vaccine misinformation affects intentions in controlled settings, but does not link it to real-world vaccination rates. Here, we present observational evidence of a causal relationship between exposure to antivaccine content and vaccination rates, and estimate the size of this effect. We present a compartmental epidemic model that includes vaccination, vaccine hesitancy, and exposure to antivaccine content. We fit the model to data to determine that a geographical pattern of exposure to online antivaccine content across US counties explains reduced vaccine uptake in the same counties. We find observational evidence that exposure to antivaccine content on Twitter caused about 14,000 people to refuse vaccination between February and August 2021 in the US, resulting in at least 545 additional cases and 8 additional deaths. This work provides a methodology for linking online speech with offline epidemic outcomes. Our findings should inform social media moderation policy as well as public health interventions.
}


\maketitle

\section{Introduction}

Vaccines were critical in reducing hospitalizations and mortality during the COVID-19 pandemic~\cite{Suthare069317,gupta2021vaccinations,schneider2021hospitalizations}. Despite their wide availability in the United States, 62\% of Americans chose not to be vaccinated during 2021~\cite{mathieu2021global}. 
Several socio-economic factors have been linked to vaccine hesitancy using correlational studies~\cite{yasminCOVID19VaccineHesitancy2021}, but finding evidence of drivers remains a challenge. One potential driver is exposure to anti-vaccine content on social media~\cite{DeVerna2024info-epidemic}. 
A laboratory study demonstrated that exposure to COVID-19 misinformation decreased willingness to be vaccinated~\cite{loomba2021measuring}. Using social media data, higher rates of misinformation were found to precede increases in COVID-19 infections in countries during early 2020~\cite{Gallotti2020,steven_lloyd_wilson_social_2020}, though this may have been due to a general wave of COVID-19 discussion in early 2020~\cite{yangthecopvid2021}. A temporal correlation was also found between increased production of online vaccine misinformation and higher vaccine hesitancy, as well as lower uptake, across US states and counties~\cite{pierri2022online}. A recent study combined experimental measurements of headlines' effects on vaccination intentions with data on URL views to explore the causal link between exposure to vaccine-skeptical content on Facebook and vaccine hesitancy~\cite{Allen24Science}. Many factors influence disease perceptions and vaccination decisions~\cite{DEGAETANO2024109337}, so it is unclear how intention to vaccinate after exposure to online content translates into actual vaccine uptake.

To establish a causal link between real-world exposure to antivaccine content and vaccine uptake rates, we must begin by measuring these two specific variables. Here we measure the extent to which changes in levels of exposure to antivaccine content on Twitter (now X) in US counties caused changes not only in vaccine hesitancy, but also COVID-19 vaccination rates and cases. 
To estimate the causal link between exposure to antivaccine tweets, vaccine hesitancy, reduced vaccine uptake, and cases, we develop an original extension to the SIR epidemic model with states that represent vaccinated and vaccine-hesitant people. This model fits empirical data about COVID cases, vaccinations, and exposure to antivaccine tweets better than simpler models that ignore vaccine hesitancy. A key parameter is the rate at which people exposed to antivaccine content become vaccine hesitant. Fitting the model to the data yields a positive value for this parameter, indicating that increased exposure decreases vaccine uptake rates.

\begin{figure}[t]
\centering
  \includegraphics[width=0.75\columnwidth]{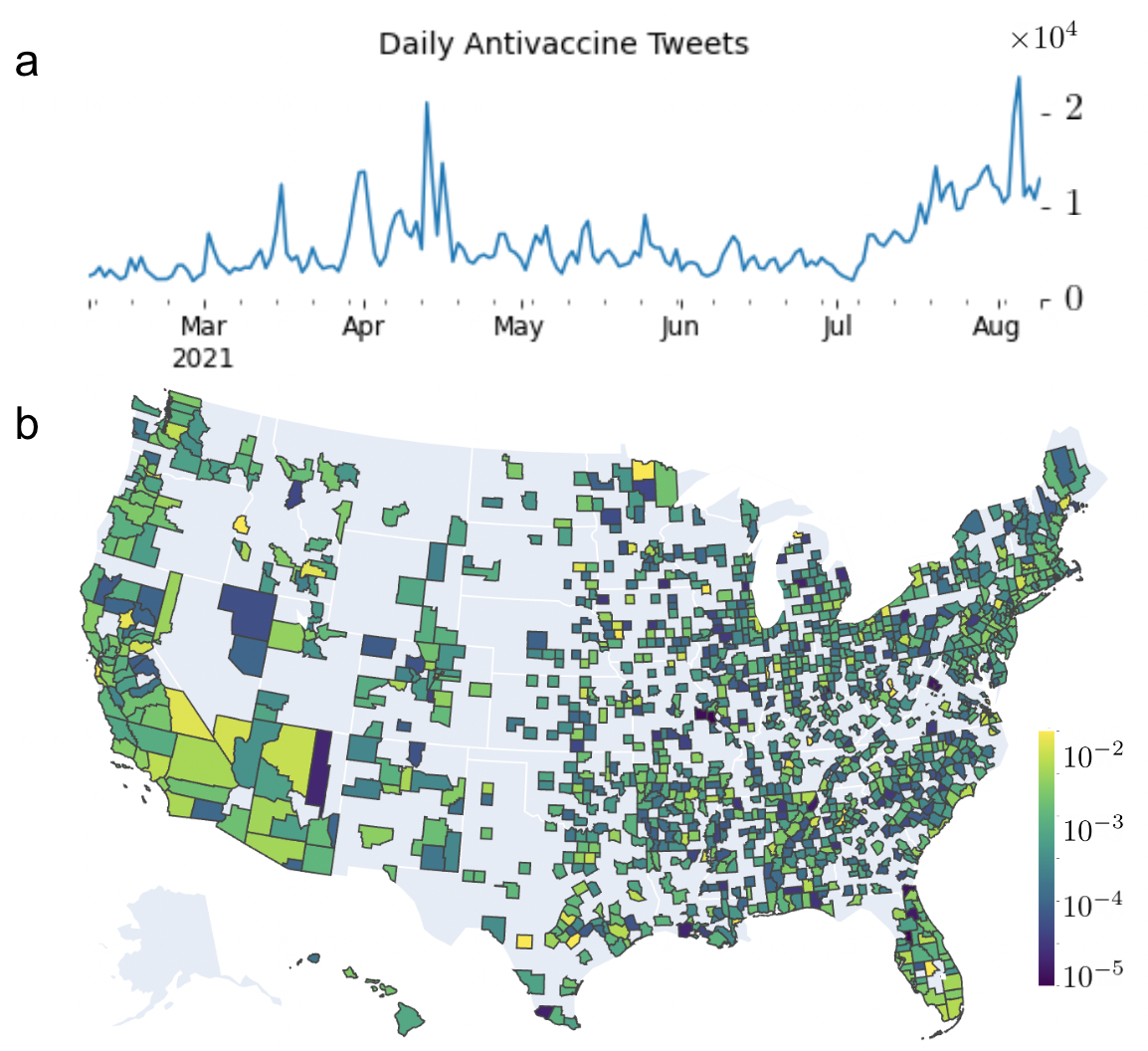}
\caption{\textbf{Antivaccine tweets.} (a) Number of geolocated antivaccine tweets for each day of the observation period. (b) Antivaccine tweets per capita per day, geolocated in each US county during the observation period. Grey coloring denotes counties where we have insufficient geolocated Twitter data.}
\label{fig:antivax_tweets_prevalence} 
\end{figure}

We analyze the records of cases and vaccinations in US counties between February and August 2021 (see Methods). To connect these data to antivaccine content exposure, we identify geolocated antivaccine tweets. 
We use a text classifier (see Methods) to identify COVID-related tweets as ``Antivax'' or ``Other.'' About 10\% of all the tweets in our data set can be geolocated to specific US counties, and about 8\% of the geolocated tweets are identified as containing antivaccine content. This yields a dataset of 26 million geolocated tweets, 2.2 million of which are identified as containing antivaccine content. Figure~\ref{fig:antivax_tweets_prevalence} shows that antivaccine content increased around July 4th, 2021 and was broadly distributed throughout the US with limited geographic clustering; some counties produced orders of magnitude more antivaccine content per capita than others. To measure county-level exposure to antivaccine content, we combine the number of antivaccine tweets produced in each county with a network that captures the spread of this content from one county to another via retweets (see Methods). 

We combine this antivaccine content exposure data with COVID case and vaccination data for each county. Next we describe the model and fit its parameters to this data, allowing us to infer the effect of antivaccine tweets on vaccine hesitancy, vaccinations, cases, and deaths. 
    
\section{Antivaccine Tweets Increase Vaccine Hesitancy}

\begin{figure}
\centering
  \includegraphics[width=1\columnwidth]{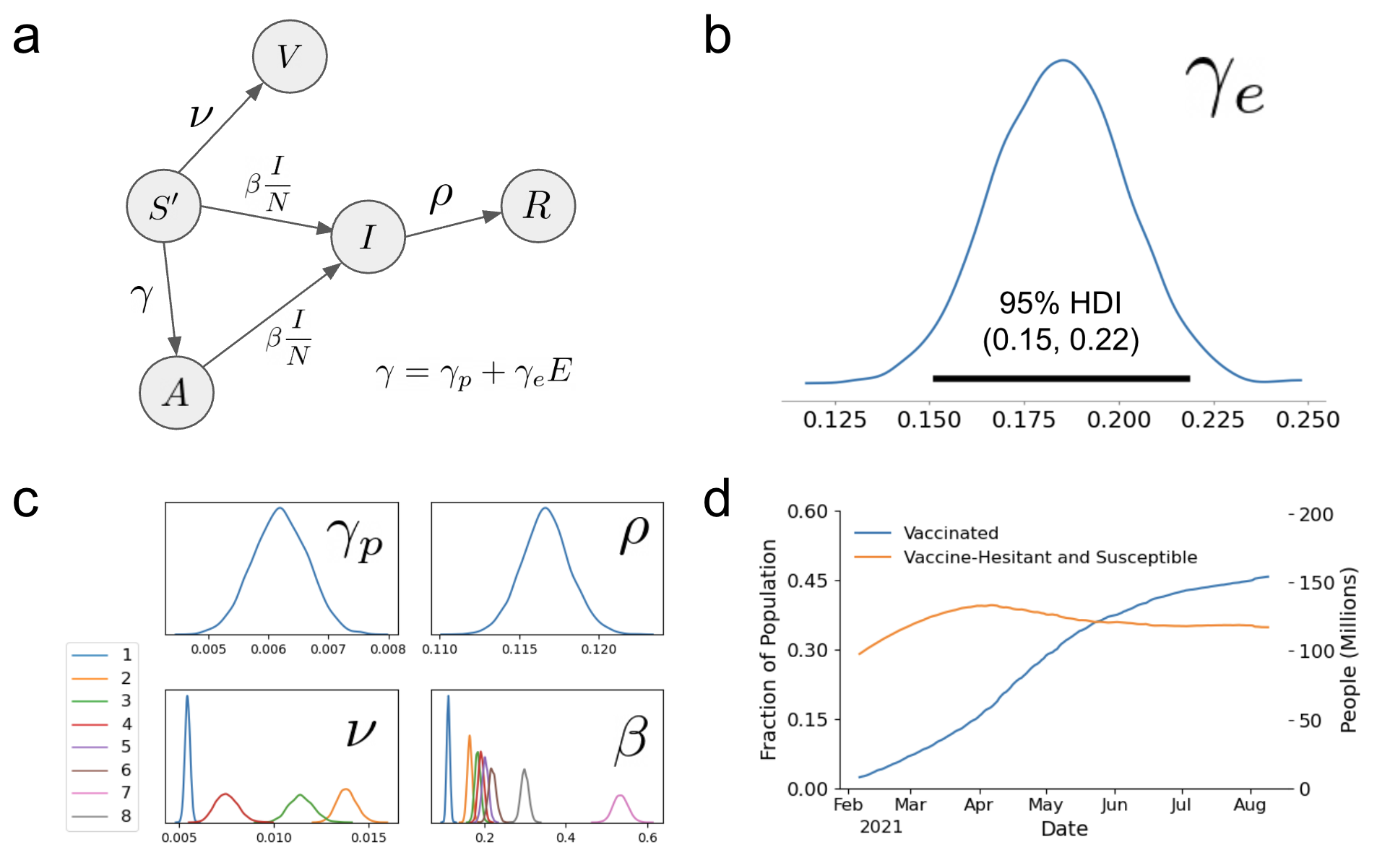}
\caption{\textbf{SIRVA model.} (a) Compartmental model diagram (see Methods). Note that $A=\alpha S$ and $S'=(1-\alpha)S$, where $\alpha$ is the vaccine hesitancy ratio, i.e., the fraction of susceptibles who are unwilling to be vaccinated at time $t$. $E$ is the magnitude of exposure to antivaccine tweets and $\gamma_e$ is the rate at which people become vaccine-hesitant due to this exposure.  (b) Posterior distribution of $\gamma_e$, with 95\% high-density interval. The posterior mean value is $\gamma_e \approx 0.18$. (c)  Posterior distributions of other global model parameters. Note that there are multiple curves for $\nu$ and $\beta$ because we used different values of these parameters during different time periods to account for changing infectivity and national vaccine availability (24-day periods for $\beta$, 48-day periods for $\nu$); these are ordered from earliest to latest. (d) Vaccinated population (V), and population who are both susceptible and vaccine-hesitant (A).}
\label{fig:model_and_posteriors}
\end{figure}

To infer the impact of antivaccine tweets on vaccine hesitancy, we model the COVID epidemic with an SIR-like compartmental model that we call SIRVA (see Methods). In addition to the standard Infected ($I$) and Recovered ($R$) compartments, the model has compartments for vaccinated people ($V$) and divides the Susceptible group ($S$) into those who are willing ($S'$) and unwilling ($A$) to be vaccinated, see Figure~\ref{fig:model_and_posteriors}a. The key epidemic model parameters are the infection rate $\beta$, the recovery rate $\rho$, the vaccination rate $\nu$, and the rate $\gamma$ at which people become unwilling to be vaccinated. The latter can be written as $\gamma = \gamma_e E + \gamma_p$, where $\gamma_e$ is the rate at which people become vaccine-hesitant per unit of exposure to antivaccine content ($E$, see Methods) and $\gamma_p$ is the rate of conversion to vaccine hesitancy due to other factors. Finally, we express the vaccine-hesitant population at time $t$ as $A=\alpha_t S$, where the vaccine hesitancy ratio $\alpha_t$ changes by rate $\gamma$ each day and its initial value is an additional parameter $\alpha_0$.

\newpage

We apply this model to each county and use Bayesian Markov Chain Monte Carlo (MCMC) to infer the posterior distributions of the parameters from the data (see Methods). We are primarily interested in the parameter $\gamma_e$, which quantifies the impact of exposure on vaccine hesitancy. Inspecting the posterior distributions of the parameters (see Figure~\ref{fig:model_and_posteriors}b,c), we find that $\gamma_e$ is greater than zero ($p = 0.0002$) with an approximate magnitude of 0.18 and a 95\% credible interval between 0.15 and 0.22. This indicates that increases in exposure to antivaccine content predict future increases in vaccine hesitancy, and subsequent decreases in the vaccine uptake rate. 

\section{Antivaccine Tweets Prevent Vaccinations}

The SIRVA model does not specify a direct relationship between exposure and vaccine uptake rates, however we can use causal graphical modeling~\cite{pearlDoCalculusRevisited2012} to assess this relationship (see Methods). 

We define the Average Treatment Effect (ATE) as the change in vaccinations per capita per unit exposure to antivaccine tweets, where exposure is expressed in units of antivaccine tweets per capita. We derive an expression for the ATE and find its magnitude to be $-3.2 \times 10^{-4}$ (see Methods). Note that the confidence that the ATE is less than zero is the same as the confidence that $\gamma_e$ is greater than zero ($p = 0.0002$). Based on this ATE, we estimate that antivaccine tweets induced 14,086 people to refuse COVID vaccinations nationwide (with a 95\% credible interval of 11,414 -- 16,759) during the period from February 6th to August 9th, 2021 in the United States (see Methods). This represents only a small fraction of the total number of Americans who were unwilling to be vaccinated and were susceptible to infection ($A$), which we estimate at approximately 117 million people in August 2021, up from 98 million in February (see Figure~\ref{fig:model_and_posteriors}d and Methods).

We can use measurements of vaccine effectiveness to provide a lower bound for the number of COVID cases and deaths that may have resulted from this reduction in vaccination. We estimate that among the 14,086 people who remained unvaccinated as a result of antivaccine tweets, there were about 545 COVID cases and 8 COVID-attributable deaths during February-August 2021 that would have been prevented without this additional vaccine hesitancy (see Methods). This represents a lower bound on the total impact because there would have been secondary infections outside the vaccine-hesitant population. Additionally, there would have been more cases and deaths following August 2021, which are not counted here.

To test if accounting for vaccine hesitancy improves model accuracy, we compared the predictions of our SIRVA model against a simpler model that does not include vaccine hesitancy (called SIRV, see Methods) using leave-one-out cross-validation and a Bayesian model fit score (see Methods). 
The standard errors of the score give a natural scale for comparing the relative accuracy of the models. 
The SIRVA model's fit score is three standard errors better than that of SIRV ($-(1716 \pm 7) \times 10^2$ vs. $-(1737 \pm 7) \times 10^2$, closer to zero is better). This indicates that SIRVA is more accurate at predicting unobserved data points.

Coupled with our consideration of potential confounding factors (see Discussion), these results allow us to infer that the observed relationship between exposure antivaccine tweets and vaccine refusal is causal. We also compare our results to a simpler linear model relating exposure to vaccine uptake, and find a similar (but stronger) negative relationship; our method improves over the simpler linear model by accounting for additional causal confounders (see Discussion).

\section{Discussion}

The proposed SIRVA model provides us with a novel approach to capture the role of vaccine hesitancy in epidemics. We use this model to measure the effect of antivaccine Twitter content on vaccine uptake during the COVID-19 pandemic in the United States. We find evidence that exposure to antivaccine content on Twitter caused decreased vaccine uptake rates and increased cases and deaths. While this analysis was based on COVID, it may have implications for outbreaks of other vaccine-preventable diseases, such as measles.

Our causal analysis hinges on a few key points. First and most critically, by leveraging the retweet network of COVID-related tweets between US counties, our measure of exposure to antivaccine content is specific to both the Twitter platform and the particular geographic distribution of Twitter users discussing COVID, allowing us to rule out many possible confounding factors that act through other social networks (e.g., Facebook) or with different geographic distributions. We test this geographic and platform specificity of our measure of exposure in two ways: (i) we tested the correlation of the COVID-related retweet network
with other social networks (i.e., Meta's Social Connectedness Index~\cite{social_connectedness_index_Bailey2018}), and found no correlation; and (ii) we tested whether shuffling exposure data by county would destroy the measured relationship between antivaccine tweets and increased vaccine hesitancy, and found that the relationship was null in this case. 

In addition to this platform and geographic specificity, we accounted for other possible causal confounders in our model, including preexisting vaccine hesitancy in each county, nationwide drift in vaccine hesitancy over time, and possible differential antivaccine content exposure rates based on preexisting vaccine hesitancy. We accounted for pre-existing antivaccine sentiments in each county by including an inferred free parameter for the initial antivaccine hesitancy ratio $\alpha_0$ in each county; this accounts for the impact that demographic and pre-existing political factors may have on antivaccine sentiment. We accounted for nationwide drift in vaccine hesitancy over time through an additional parameter $\gamma_p$ allowing for conversion to vaccine hesitancy without exposure to antivaccine tweets. Additionally, in our causal graphical model, we accounted for the tendency of people with higher vaccine hesitancy to be more likely to be exposed to antivaccine content on social media. To this end, we included an additional causal path from vaccine hesitancy to exposure, and this is reflected in our average treatment effect estimation. This explicitly accounts for any possible ``reverse causation'' feedback from vaccine hesitancy to exposure. 

We also tested an alternative SIRVA model in which existing vaccine hesitancy may produce additional vaccine hesitancy within a county, for instance by word-of-mouth spread of antivaccine sentiment (see Methods); this more complex model produced qualitatively similar results as the SIRVA model, ruling out the possibility that word-of-mouth spread can explain the relationships we find. Between testing this alternative model and including the feedback from vaccine hesitancy to exposure in our causal graph, we believe we have accounted for any endogenous dynamics that could confound our measurement.

Although our causal inference analysis accounts for confounding factors that affect the vaccine uptake rate by inducing vaccine hesitancy, we have not accounted for confounders that might act via vaccine availability. We assume that the processes governing Twitter's social dynamics and those determining vaccine availability are largely independent. Based on this assumption, we believe there is no significant common cause of both vaccine unavailability and exposure to antivaccine tweets. Because effects on vaccine uptake must act either by vaccine hesitancy or vaccine availability, we believe we have accounted for the possible confounders. We therefore conclude that the observed relationship between exposure to antivaccine tweets and reduced vaccine uptake rate is causal.

There are different ways to model the effects of antivaccine content exposure on vaccination uptake. For instance, we could alternatively make the daily vaccination rate ($\nu$) dependent on exposure ($E$). 
This would be simpler than the SIRVA model, by eliminating the compartment A and the latent variables that are used to construct it. 
However, the SIRVA model lets us capture the intuition that vaccine hesitancy is a durable result of accumulated exposures over time rather than an immediate and transient effect of the most recent media exposures.

The accuracy of our method may be affected by data limitations and modeling assumptions. First, the official CDC data on COVID vaccinations, cases, and deaths contain imputed values and reporting lags. Second, our Twitter user geolocation data has limited coverage of the user population. Third, our observations may not capture the full spectrum of all antivaccine content on Twitter. These limitations may have created sample biases in the measured antivaccine content exposure and vaccine uptake rates. 

On the modeling side, we neglect potential infections among vaccinated people, vaccinations among vaccine-hesitant people, and the lag time between exposure to COVID and active infection. In addition, while variations of epidemic models have been proposed to capture the spread of misinformation (e.g., \cite{Jin2013Epidemiological,Kauk2021Understanding,Pilditch2022Psychological}), we simply assume that the flow of antivaccine content between counties is proportional to retweet rates. Finally, we chose to measure an average treatment effect over all counties together, using a single key parameter $\gamma_e$ to capture the effect size, rather than attempting to measure the effect in each county individually or in smaller groupings of counties; this averaging was necessary to resolve a clear signal in our measurement, but makes the effect size seem homogeneous across all counties, when in fact it likely differs between counties. Future work could explore the heterogeneity of $\gamma_e$ across different (subsets of) counties. For example, one might expect different effects ($\gamma_e$ values) based on the political leanings of counties~\cite{pierri2022online}. 

With these caveats, our analysis estimates that 14,086 people became vaccine hesitant as a result of antivaccine content on Twitter. These are only a very small fraction of our model's estimate of 27 million Americans who became vaccine hesitant between February and August 2021. This larger increase in hesitancy is likely due to other sources of antivaccine messaging outside Twitter, including Facebook~\cite{Allen24Science}, traditional media, and word-of-mouth interactions. Further work could analyze data from other platforms, such as Instagram and TikTok, which also carry antivaccination content~\cite{ortiz-sanchez_analysis_2020}.

The impact of antivaccine content on public health outcomes highlighted by our analysis suggests that policy countermeasures may be necessary. Prior literature suggests a few possible approaches. Public health information campaigns may be used to ``inoculate'' the public against misinformation by spreading true information first~\cite{Roozenbeek2022Psychological,Lu2024Psychological,Pilditch2022Psychological}, or to actively fact check misinformation that is already spreading~\cite{Kreps2022Infodemic,Bowles2025Sustaining,Kauk2021Understanding}. Limiting the spread of misinformation on social media by removing posts or reducing their visibility can also be effective at reducing their impact~\cite{Butts2023Mathematical,Kauk2021Understanding}, especially if combined with other methods~\cite{Bak-Coleman2022Combining}. Similarly, removing ``superspreaders'' who propagate a large fraction of the misinformation on a platform can be an effective intervention~\cite{DeVerna_2024}. In general, earlier interventions seem to work better than later ones~\cite{Brashier2021TimingMatters}.

This work contributes to both public health research and social media studies by establishing a causal link between online content and offline public health outcomes. These conclusions should inform future social media policy and epidemic modeling efforts.

\section{Methods}
\label{sec11}

\subsection{Data availability}

We use three main data sources in this work. The CoVaxxy tweets database \cite{covaxxy} contains tweets related to COVID vaccines from January 2021 to February 2023 and is available at \url{doi.org/10.5281/zenodo.4526494}.
In this study we used tweets geolocated to US counties from February 6th to August 9th, 2021. The beginning of this window coincides with vaccines becoming widely available in the US, and the end of the window marks approximately the time when the vaccine uptake rate began to slow substantially. The primary features of this data are the text content, timestamps, and geolocations of the tweets. 
In addition, we used CDC records of COVID cases, deaths, and vaccinations in US counties \cite{CDC2021}, and Mønsted and Lehmann's antivaccine tweets dataset \cite{monstedAlgorithmicDetectionAnalysis2019}, which supplements our own labeled data for training a tweet classifier. 
Additional details are provided in Supplementary Methods. Code is available at \url{github.com/osome-iu/effects_of_antivax_tweets_on_covid19_outcomes}.

\subsection{Antivaccine Tweet Classifier} \label{sec:antivax_classifier}

To track the volume of antivaccine content produced in each county, we built a text classifier that determines if a tweet expresses antivaccine sentiment. The classifier takes the text of a tweet as input and returns a label, either ``antivaccine'' or ``other.''  It was trained on 4,200 tweets labeled by two human annotators and 2,000 tweets labeled by Mønsted and Lehmann~\cite{monstedAlgorithmicDetectionAnalysis2019}. The classifier achieved good accuracy ($F_1=0.74$) on a hold-out set of 900 tweets, which were randomly sampled from our set of geolocated covid-related tweets. Further details about the data annotations, classification model, and results are found in Supplementary Methods.

\subsection{Exposure to Antivaccine Tweets} \label{sec:exposure_definition}

To measure the impact of antivaccine tweets on people's propensity to get vaccinated, we measure the amount of antivaccine Twitter content to which people are exposed at the county level. Intuitively, if people in county $i$ reshare a lot of content from people in county $j$, this implies that they are exposed to content from county $j$ --- exposure is a necessary, though not sufficient, condition for sharing. Therefore we can use resharing of COVID-related content as proxy for exposure to such content. We further assume that exposure to anti-vaccine content is proportional to exposure to COVID-related content in general. Based on these assumptions, the per-capita antivaccine exposure rate in county $i$ at time $t$ is defined as
\begin{align}\label{eq:exposure}
    E_{i,t} = \frac{1}{N_i} \frac{\sum_j W_{ij} T_{j,t}}{\sum_j W_{ij}}
\end{align}
where $N_i$ is the population of county $i$, $T_{j,t}$ is the number of antivaccine tweets in county $j$ during time window $t$, and $W_{ij}$ is the number of times that any COVID-related tweets posted by users in county $j$ were retweeted by users in county $i$ during our observation period. Here the retweet network is only used to estimate the strength of information flow between counties; our measure of exposure does \textit{not} require that people retweet antivaccine content to be counted as exposed.

\subsection{SIRVA Model}

In the SIRVA epidemic model we developed for this work (see Figure~\ref{fig:model_and_posteriors}), we make three simplifying assumptions: first, that vaccinated people ($V$) do not become infected; second, that vaccine-hesitant people ($A=\alpha S$) do not become vaccinated; and finally, we ignore the lag time between exposure to COVID and active infection.

The dynamic equations for SIRVA in each county can be written as:
\begin{align}
    \frac{dS}{dt} &= -\beta (I/N) S - \nu S (1-\alpha) \label{eq:model-dynamics} \\
    \frac{dI}{dt} &= \beta (I/N) S - \rho I\\ 
    \frac{dR}{dt} &= \rho I \\
    \frac{dV}{dt} &= \nu S (1-\alpha) \label{eq:SIRVA_model_Vdot} \\  
    \frac{d\alpha}{dt} &= \gamma (1-\alpha) \label{eq:SIRVA_model_alphadot}
\end{align} 
where $N$ is the population of the county, $S$ is the number of people who are susceptible to infection, $I$ is the number of people who are currently infected, $R$ is the number of people who have either recovered or died from the infection, $V$ is the number of people who are vaccinated, $\alpha$ is the vaccine hesitancy ratio, $\gamma$ is the conversion rate to vaccine-hesitancy, and $\beta$, $\rho$, and $\nu$ are infection, recovery, and vaccination rate parameters, respectively. To keep the notation simple, we omit the explicit time dependency of various parameters, such as $\gamma$.

To infer the latent variable $\alpha$ from the data, we assume that there is an initial vaccine hesitancy ratio, $\alpha_0$, and people convert to vaccine hesitancy from the non-vaccine-hesitant portion of the population $(1-\alpha)$ at rate $\gamma_t$. We thus compute $\alpha_t$ as:
\begin{align}
\alpha_t = \alpha_0 + \int_{t'<t} \frac{d\alpha_{t'}}{dt'} \, dt' = \alpha_0 + \int_{t'<t} \gamma (1-\alpha_{t'}) \, dt'. \label{eq:SIRVA_model_alpha}
\end{align}
Finally, we break the conversion rate $\gamma$ into components due to exposure to antivaccine tweets ($\gamma_e$) and other factors ($\gamma_p$): $\gamma = \gamma_p + \gamma_e E_t$, where $E_t$ is the antivaccine exposure rate in the county at time $t$. For our analyses we use a time interval of 8 days between data points.

The parameter $\alpha_0$ is inferred for each county. All the other parameters ($\beta$, $\rho$, $\nu$, $\gamma_e$, and $\gamma_p$) are inferred from the data across all counties. We use Bayesian Markov Chain Monte Carlo (MCMC) sampling to infer their posterior distributions given the data (see details in Supplementary Methods).

We also define a simpler comparison model, SIRV, which is a special case of SIRVA where $\alpha_t=0$ for all $t$. 
This allows us to test how including vaccine hesitancy and its dynamics impacts the model's predictive performance. 
Additionally, we ran a static version of SIRVA, SIRVA\_static, in which the vaccine-hesitant population is static in each county, i.e., $\alpha_t = \alpha_{0}$ for all $t$, and thus $\gamma_e = \gamma_p = 0$; we found this model has slightly worse model fit than SIRVA by the ELPD-LOO criterion.

Another possible model would include a ``word of mouth'' feedback term, $\gamma_a \alpha$, in $\gamma$, accounting for the endogenous spread of vaccine hesitancy within a county as a social contagion.  We also tested a model with this dynamic effect and found similar results to the SIRVA model described here (i.e. $\gamma_e$ was of similar magnitude, as were its confidence intervals). We dropped this feature of the model for simplicity.

\subsection{Effect Estimations}

We use causal graphical modeling (see Supplementary Methods) to derive an expression for the average treatment effect (ATE) of exposure $E$ on the per capita vaccine uptake rate $\frac{1}{N_i}\dot{V}=\frac{1}{N_i}\frac{dV}{dt}$ from the dynamical equations of the SIRVA model (Eqs.~\ref{eq:model-dynamics}--\ref{eq:SIRVA_model_alpha}):
\begin{align} \label{eq:ATE_estimand_indexed}
    \text{ATE}=\frac{d}{dE_{t-1}}  \mathbb{E}\Big[\frac{\dot{V_t}}{N_i} \mid \alpha_{t-1}\Big] \approx - \mathbb{E} \Big[\frac{1}{N_i}\nu_{t-1} \, S_{t-1,i} \; \gamma_{e} \, (1-\alpha_{t-1, i}) \Big]
\end{align}
where the expectation value is taken over times $t$, counties $i$, and the posterior distributions of $\nu$, $\gamma_e$, $\alpha$, and the derivation is detailed in Supplementary Methods (Eq. A3). This results in a figure of $-3.2 \times 10^{-4}$ (with a $95\%$ credible interval between $-3.8 \times 10^{-4}$ and $-2.6 \times 10^{-4}$) vaccinations per daily tweet of exposure.

Given the ATE from Eq.~\ref{eq:ATE_estimand_indexed}, we wish to calculate how many vaccinations were prevented in each county and nationwide. Assuming the total change in vaccine uptake rate is relatively small, we can estimate the number of vaccinations per capita prevented each day in each county as
$\frac{1}{N_i}{\Delta V}_{t,i} \approx (\text{ATE})E_{t-1,i}$, 
and then use this estimation to calculate the total number of vaccines prevented nationwide as 
${\Delta V} = N \, \Delta t \; \mathbb{E}_{i,t}[\frac{\Delta V_{t,i}}{N_i}]$, where $N$ is the total population of the US and $\Delta t$ is the length of the observation period (see Supplementary Methods).

We calculate the number of people across the whole US who remained unvaccinated as a result of exposure to antivaccine tweets to be 14,086 (see Supplementary Methods). We then wish to estimate how many cases and deaths may have been prevented among these people if they had been vaccinated instead. Assuming their infection and death rates were typical for unvaccinated people during the February-August 2021 time period (3,870 cases and 57 deaths per 100,000 people, see Supplementary Methods), we estimate that about 545 COVID cases and 8 deaths that occurred in this population are attributable to antivaccine tweets, as reported in our results.  

\subsection{Model Selection Criteria} \label{sec:LOO_criterion}

We can measure each model's expected out-of-sample model performance using leave-one-out cross validation (LOO), approximated with Pareto-smoothed importance sampling (PSIS). 
The PSIS-LOO criterion~\cite{vehtariPracticalBayesianModel2017} is designed to estimate out-of-sample predictive performance by approximating the expected log pointwise predictive density (ELPD) for a new dataset from the observed dataset without refitting the model to the data. The criterion is robust and efficient and represents the current state of the art for Bayesian model comparison. 
We therefore compare the performance of the SIRVA model to simpler models by measuring each model's Bayesian LOO estimate of ELPD (ELPD-LOO). Values of ELPD-LOO closer to zero indicates better fit to the data. Metrics were computed on a sampled data set of 400 random counties and 24 evenly spaced dates spanning the observation period from February to August 2021.


\backmatter

\bmhead{Abbreviations}

ATE: Average Treatment Effect. 
CDC: Centers for Disease Control and Prevention. 
CGM: Causal Graphical Model. 
ELPD: Expected Log Pointwise Predictive Density. 
HDI: High Probability Interval. 
LOO: Leave-One-Out. 
MCMC: Markov Chain Monte Carlo. 
PSIS: Pareto-Smoothed Importance Sampling. 
SIR: Susceptible-Infected-Recovered. 
SIRV: Susceptible-Infected-Recovered-Vaccinated. 
SIRVA: Susceptible-Infected-Recovered-Vaccinated-Antivaccine. 

\bmhead{Supplementary information}

Supplementary Methods are included in Appendix.

\section*{Declarations}

\subsection*{Ethics approval}

The collection of public posts from Twitter was deemed exempt from review by the Indiana University Institutional Review Board (protocol 1102004860).

\subsection*{Data and code availability}

We use three main data sources in this work: the CoVaxxy Tweets database, which contains Tweets related to COVID vaccines \cite{covaxxy} and is available at \url{doi.org/10.5281/zenodo.4526494}; CDC records of COVID cases, deaths, and vaccinations in US counties \cite{CDC2021}; and Mønsted and Lehmann's antivaccine tweets dataset \cite{monstedAlgorithmicDetectionAnalysis2019}, which supplements our own labeled data for training a tweet classifier. Details are provided in Supplementary Methods. Data and code are available at \url{github.com/osome-iu/effects_of_antivax_tweets_on_covid19_outcomes}. 

\subsection*{Funding}

This work was supported in part by the Knight Foundation and Craig Newmark Philanthropies.

\subsection*{Author contribution}

All authors contributed to conceptualization of the work and data acquisition. JBr and Jbo developed the methodology. JBo led the investigation (software development, data curation and analysis, results validation, visualization) and wrote the original draft. JBr and FM provided supervision and reviewed and edited the manuscript. FM acquired lab funding and administered the project.

\subsection*{Acknowledgements}

We are grateful to Marissa Donofrio for annotating tweets, to Bjarke M{\o}nsted and Sune Lehmann for sharing additional annotation data, and to Vincent Jansen, Alessandro Flammini, and Yong-Yeol Ahn for helpful discussion. We also thank four anonymous reviewers who provided helpful feedback, highlighting an error in an earlier version of the manuscript.

\newpage
\bibliography{main}


\begin{thebibliography}{44}
\ifx \bisbn   \undefined \def \bisbn  #1{ISBN #1}\fi
\ifx \binits  \undefined \def \binits#1{#1}\fi
\ifx \bauthor  \undefined \def \bauthor#1{#1}\fi
\ifx \batitle  \undefined \def \batitle#1{#1}\fi
\ifx \bjtitle  \undefined \def \bjtitle#1{#1}\fi
\ifx \bvolume  \undefined \def \bvolume#1{\textbf{#1}}\fi
\ifx \byear  \undefined \def \byear#1{#1}\fi
\ifx \bissue  \undefined \def \bissue#1{#1}\fi
\ifx \bfpage  \undefined \def \bfpage#1{#1}\fi
\ifx \blpage  \undefined \def \blpage #1{#1}\fi
\ifx \burl  \undefined \def \burl#1{\textsf{#1}}\fi
\ifx \doiurl  \undefined \def \doiurl#1{\url{https://doi.org/#1}}\fi
\ifx \betal  \undefined \def \betal{\textit{et al.}}\fi
\ifx \binstitute  \undefined \def \binstitute#1{#1}\fi
\ifx \binstitutionaled  \undefined \def \binstitutionaled#1{#1}\fi
\ifx \bctitle  \undefined \def \bctitle#1{#1}\fi
\ifx \beditor  \undefined \def \beditor#1{#1}\fi
\ifx \bpublisher  \undefined \def \bpublisher#1{#1}\fi
\ifx \bbtitle  \undefined \def \bbtitle#1{#1}\fi
\ifx \bedition  \undefined \def \bedition#1{#1}\fi
\ifx \bseriesno  \undefined \def \bseriesno#1{#1}\fi
\ifx \blocation  \undefined \def \blocation#1{#1}\fi
\ifx \bsertitle  \undefined \def \bsertitle#1{#1}\fi
\ifx \bsnm \undefined \def \bsnm#1{#1}\fi
\ifx \bsuffix \undefined \def \bsuffix#1{#1}\fi
\ifx \bparticle \undefined \def \bparticle#1{#1}\fi
\ifx \barticle \undefined \def \barticle#1{#1}\fi
\bibcommenthead
\ifx \bconfdate \undefined \def \bconfdate #1{#1}\fi
\ifx \botherref \undefined \def \botherref #1{#1}\fi
\ifx \url \undefined \def \url#1{\textsf{#1}}\fi
\ifx \bchapter \undefined \def \bchapter#1{#1}\fi
\ifx \bbook \undefined \def \bbook#1{#1}\fi
\ifx \bcomment \undefined \def \bcomment#1{#1}\fi
\ifx \oauthor \undefined \def \oauthor#1{#1}\fi
\ifx \citeauthoryear \undefined \def \citeauthoryear#1{#1}\fi
\ifx \endbibitem  \undefined \def \endbibitem {}\fi
\ifx \bconflocation  \undefined \def \bconflocation#1{#1}\fi
\ifx \arxivurl  \undefined \def \arxivurl#1{\textsf{#1}}\fi
\csname PreBibitemsHook\endcsname

\bibitem[\protect\citeauthoryear{Suthar et~al.}{2022}]{Suthare069317}
\begin{botherref}
\oauthor{\bsnm{Suthar}, \binits{A.B.}},
\oauthor{\bsnm{Wang}, \binits{J.}},
\oauthor{\bsnm{Seffren}, \binits{V.}},
\oauthor{\bsnm{Wiegand}, \binits{R.E.}},
\oauthor{\bsnm{Griffing}, \binits{S.}},
\oauthor{\bsnm{Zell}, \binits{E.}}:
{Public health impact of covid-19 vaccines in the US: observational study}.
BMJ
\textbf{377}
(2022)
\doiurl{10.1136/bmj-2021-069317}
\end{botherref}
\endbibitem

\bibitem[\protect\citeauthoryear{Gupta et~al.}{2021}]{gupta2021vaccinations}
\begin{barticle}
\bauthor{\bsnm{Gupta}, \binits{S.}},
\bauthor{\bsnm{Cantor}, \binits{J.}},
\bauthor{\bsnm{Simon}, \binits{K.I.}},
\bauthor{\bsnm{Bento}, \binits{A.I.}},
\bauthor{\bsnm{Wing}, \binits{C.}},
\bauthor{\bsnm{Whaley}, \binits{C.M.}}:
\batitle{{Vaccinations Against COVID-19 May Have Averted Up To 140,000 Deaths
  In The United States}}.
\bjtitle{Health Affairs}
\bvolume{40}(\bissue{9}),
\bfpage{1465}--\blpage{1472}
(\byear{2021})
\doiurl{10.1377/hlthaff.2021.00619}
\end{barticle}
\endbibitem

\bibitem[\protect\citeauthoryear{Schneider
  et~al.}{2021}]{schneider2021hospitalizations}
\begin{botherref}
\oauthor{\bsnm{Schneider}, \binits{E.C.}},
\oauthor{\bsnm{Shah}, \binits{A.}},
\oauthor{\bsnm{Sah}, \binits{P.}},
\oauthor{\bsnm{Moghadas}, \binits{S.M.}},
\oauthor{\bsnm{Vilches}, \binits{T.}},
\oauthor{\bsnm{Galvani}, \binits{A.}}:
{The U.S. COVID-19 Vaccination Program at One Year: How Many Deaths and
  Hospitalizations Were Averted?}
(2021).
\url{https://www.commonwealthfund.org/publications/issue-briefs/2021/dec/us-covid-19-vaccination-program-one-year-how-many-deaths-and}
Accessed 2021-12-14
\end{botherref}
\endbibitem

\bibitem[\protect\citeauthoryear{Mathieu et~al.}{2021}]{mathieu2021global}
\begin{barticle}
\bauthor{\bsnm{Mathieu}, \binits{E.}},
\bauthor{\bsnm{Ritchie}, \binits{H.}},
\bauthor{\bsnm{Ortiz-Ospina}, \binits{E.}},
\bauthor{\bsnm{Roser}, \binits{M.}},
\bauthor{\bsnm{Hasell}, \binits{J.}},
\bauthor{\bsnm{Appel}, \binits{C.}},
\bauthor{\bsnm{Giattino}, \binits{C.}},
\bauthor{\bsnm{Rod{\'e}s-Guirao}, \binits{L.}}:
\batitle{{A global database of COVID-19 vaccinations}}.
\bjtitle{Nature human behaviour}
\bvolume{5}(\bissue{7}),
\bfpage{947}--\blpage{953}
(\byear{2021})
\doiurl{10.1038/s41562-021-01122-8}
\end{barticle}
\endbibitem

\bibitem[\protect\citeauthoryear{Yasmin
  et~al.}{2021}]{yasminCOVID19VaccineHesitancy2021}
\begin{barticle}
\bauthor{\bsnm{Yasmin}, \binits{F.}},
\bauthor{\bsnm{Najeeb}, \binits{H.}},
\bauthor{\bsnm{Moeed}, \binits{A.}},
\bauthor{\bsnm{Naeem}, \binits{U.}},
\bauthor{\bsnm{Asghar}, \binits{M.S.}},
\bauthor{\bsnm{Chughtai}, \binits{N.U.}},
\bauthor{\bsnm{Yousaf}, \binits{Z.}},
\bauthor{\bsnm{Seboka}, \binits{B.T.}},
\bauthor{\bsnm{Ullah}, \binits{I.}},
\bauthor{\bsnm{Lin}, \binits{C.-Y.}},
\bauthor{\bsnm{Pakpour}, \binits{A.H.}}:
\batitle{{{COVID-19 Vaccine Hesitancy}} in the {{United States}}: {{A
  Systematic Review}}}.
\bjtitle{Frontiers in Public Health}
\bvolume{9},
\bfpage{770985}
(\byear{2021})
\doiurl{10.3389/fpubh.2021.770985}
\end{barticle}
\endbibitem

\bibitem[\protect\citeauthoryear{DeVerna
  et~al.}{2025}]{DeVerna2024info-epidemic}
\begin{barticle}
\bauthor{\bsnm{DeVerna}, \binits{M.R.}},
\bauthor{\bsnm{Pierri}, \binits{F.}},
\bauthor{\bsnm{Ahn}, \binits{Y.-Y.}},
\bauthor{\bsnm{Fortunato}, \binits{S.}},
\bauthor{\bsnm{Flammini}, \binits{A.}},
\bauthor{\bsnm{Menczer}, \binits{F.}}:
\batitle{Modeling the amplification of epidemic spread by individuals exposed
  to misinformation on social media}.
\bjtitle{npj Complexity}
\bvolume{2},
\bfpage{11}
(\byear{2025})
\doiurl{10.1038/s44260-025-00038-y}
\end{barticle}
\endbibitem

\bibitem[\protect\citeauthoryear{Loomba et~al.}{2021}]{loomba2021measuring}
\begin{barticle}
\bauthor{\bsnm{Loomba}, \binits{S.}},
\bauthor{\bsnm{Figueiredo}, \binits{A.}},
\bauthor{\bsnm{Piatek}, \binits{S.J.}},
\bauthor{\bsnm{Graaf}, \binits{K.}},
\bauthor{\bsnm{Larson}, \binits{H.J.}}:
\batitle{{Measuring the impact of COVID-19 vaccine misinformation on
  vaccination intent in the UK and USA}}.
\bjtitle{Nature human behaviour}
\bvolume{5}(\bissue{3}),
\bfpage{337}--\blpage{348}
(\byear{2021})
\doiurl{10.1038/s41562-021-01056-1}
\end{barticle}
\endbibitem

\bibitem[\protect\citeauthoryear{Gallotti et~al.}{2020}]{Gallotti2020}
\begin{barticle}
\bauthor{\bsnm{Gallotti}, \binits{R.}},
\bauthor{\bsnm{Valle}, \binits{F.}},
\bauthor{\bsnm{Castaldo}, \binits{N.}},
\bauthor{\bsnm{Sacco}, \binits{P.}},
\bauthor{\bsnm{De~Domenico}, \binits{M.}}:
\batitle{{Assessing the risks of `infodemics' in response to COVID-19
  epidemics}}.
\bjtitle{Nature Human Behaviour}
\bvolume{4}(\bissue{12}),
\bfpage{1285}--\blpage{1293}
(\byear{2020})
\doiurl{10.1038/s41562-020-00994-6}
\end{barticle}
\endbibitem

\bibitem[\protect\citeauthoryear{{Steven Lloyd Wilson} and {Charles
  Wiysonge}}{2020}]{steven_lloyd_wilson_social_2020}
\begin{barticle}
\bauthor{\bsnm{{Steven Lloyd Wilson}}},
\bauthor{\bsnm{{Charles Wiysonge}}}:
\batitle{Social media and vaccine hesitancy}.
\bjtitle{BMJ Global Health}
\bvolume{5}(\bissue{10}),
\bfpage{004206}
(\byear{2020})
\doiurl{10.1136/bmjgh-2020-004206}
\end{barticle}
\endbibitem

\bibitem[\protect\citeauthoryear{Yang et~al.}{2021}]{yangthecopvid2021}
\begin{barticle}
\bauthor{\bsnm{Yang}, \binits{K.-C.}},
\bauthor{\bsnm{Pierri}, \binits{F.}},
\bauthor{\bsnm{Hui}, \binits{P.-M.}},
\bauthor{\bsnm{Axelrod}, \binits{D.}},
\bauthor{\bsnm{Torres-Lugo}, \binits{C.}},
\bauthor{\bsnm{Bryden}, \binits{J.}},
\bauthor{\bsnm{Menczer}, \binits{F.}}:
\batitle{The covid-19 infodemic: Twitter versus facebook}.
\bjtitle{Big Data \& Society}
\bvolume{8}(\bissue{1}),
\bfpage{20539517211013861}
(\byear{2021})
\doiurl{10.1177/20539517211013861}
\end{barticle}
\endbibitem

\bibitem[\protect\citeauthoryear{Pierri et~al.}{2022}]{pierri2022online}
\begin{barticle}
\bauthor{\bsnm{Pierri}, \binits{F.}},
\bauthor{\bsnm{Perry}, \binits{B.L.}},
\bauthor{\bsnm{DeVerna}, \binits{M.R.}},
\bauthor{\bsnm{Yang}, \binits{K.-C.}},
\bauthor{\bsnm{Flammini}, \binits{A.}},
\bauthor{\bsnm{Menczer}, \binits{F.}},
\bauthor{\bsnm{Bryden}, \binits{J.}}:
\batitle{{Online misinformation is linked to early COVID-19 vaccination
  hesitancy and refusal}}.
\bjtitle{Scientific Reports}
\bvolume{12}(\bissue{1}),
\bfpage{1}--\blpage{7}
(\byear{2022})
\doiurl{10.1038/s41598-022-10070-w}
\end{barticle}
\endbibitem

\bibitem[\protect\citeauthoryear{Allen et~al.}{2024}]{Allen24Science}
\begin{barticle}
\bauthor{\bsnm{Allen}, \binits{J.}},
\bauthor{\bsnm{Watts}, \binits{D.J.}},
\bauthor{\bsnm{Rand}, \binits{D.G.}}:
\batitle{Quantifying the impact of misinformation and vaccine-skeptical content
  on facebook}.
\bjtitle{Science}
\bvolume{384}(\bissue{6699}),
\bfpage{3451}
(\byear{2024})
\doiurl{10.1126/science.adk3451}
\end{barticle}
\endbibitem

\bibitem[\protect\citeauthoryear{{De Gaetano}
  et~al.}{2024}]{DEGAETANO2024109337}
\begin{barticle}
\bauthor{\bsnm{{De Gaetano}}, \binits{A.}},
\bauthor{\bsnm{Barrat}, \binits{A.}},
\bauthor{\bsnm{Paolotti}, \binits{D.}}:
\batitle{Modeling the interplay between disease spread, behaviors, and disease
  perception with a data-driven approach}.
\bjtitle{Mathematical Biosciences}
\bvolume{378},
\bfpage{109337}
(\byear{2024})
\doiurl{10.1016/j.mbs.2024.109337}
\end{barticle}
\endbibitem

\bibitem[\protect\citeauthoryear{Pearl}{2012}]{pearlDoCalculusRevisited2012}
\begin{botherref}
\oauthor{\bsnm{Pearl}, \binits{J.}}:
{The Do-Calculus Revisited}.
Proc. Twenty-Eighth Conference on Uncertainty in Artificial Intelligence (UAI),
3--11
(2012)
\doiurl{10.48550/arXiv.1210.4852} .
Preprint arXiv:1210.4852
\end{botherref}
\endbibitem

\bibitem[\protect\citeauthoryear{Bailey
  et~al.}{2018}]{social_connectedness_index_Bailey2018}
\begin{barticle}
\bauthor{\bsnm{Bailey}, \binits{M.}},
\bauthor{\bsnm{Cao}, \binits{R.}},
\bauthor{\bsnm{Kuchler}, \binits{T.}},
\bauthor{\bsnm{Stroebel}, \binits{J.}},
\bauthor{\bsnm{Wong}, \binits{A.}}:
\batitle{{Social Connectedness: Measurement, Determinants, and Effects}}.
\bjtitle{Journal of Economic Perspectives}
\bvolume{32}(\bissue{3}),
\bfpage{259}--\blpage{80}
(\byear{2018})
\doiurl{10.1257/jep.32.3.259}
\end{barticle}
\endbibitem

\bibitem[\protect\citeauthoryear{Jin et~al.}{2013}]{Jin2013Epidemiological}
\begin{bchapter}
\bauthor{\bsnm{Jin}, \binits{F.}},
\bauthor{\bsnm{Dougherty}, \binits{E.}},
\bauthor{\bsnm{Saraf}, \binits{P.}},
\bauthor{\bsnm{Cao}, \binits{Y.}},
\bauthor{\bsnm{Ramakrishnan}, \binits{N.}}:
\bctitle{Epidemiological modeling of news and rumors on twitter}.
In: \bbtitle{Proceedings of the 7th Workshop on Social Network Mining and
  Analysis}.
\bsertitle{SNAKDD '13}.
\bpublisher{Association for Computing Machinery},
\blocation{New York, NY, USA}
(\byear{2013}).
\doiurl{10.1145/2501025.2501027} .
\burl{https://doi.org/10.1145/2501025.2501027}
\end{bchapter}
\endbibitem

\bibitem[\protect\citeauthoryear{Kauk et~al.}{2021}]{Kauk2021Understanding}
\begin{barticle}
\bauthor{\bsnm{Kauk}, \binits{J.}},
\bauthor{\bsnm{Kreysa}, \binits{H.}},
\bauthor{\bsnm{Schweinberger}, \binits{S.R.}}:
\batitle{Understanding and countering the spread of conspiracy theories in
  social networks: Evidence from epidemiological models of twitter data}.
\bjtitle{PLOS ONE}
\bvolume{16}(\bissue{8}),
\bfpage{0256179}
(\byear{2021})
\doiurl{10.1371/journal.pone.0256179}
\end{barticle}
\endbibitem

\bibitem[\protect\citeauthoryear{Pilditch
  et~al.}{2022}]{Pilditch2022Psychological}
\begin{barticle}
\bauthor{\bsnm{Pilditch}, \binits{T.D.}},
\bauthor{\bsnm{Roozenbeek}, \binits{J.}},
\bauthor{\bsnm{Madsen}, \binits{J.K.}},
\bauthor{\bsnm{Van Der~Linden}, \binits{S.}}:
\batitle{Psychological inoculation can reduce susceptibility to misinformation
  in large rational agent networks}.
\bjtitle{Royal Society Open Science}
\bvolume{9}(\bissue{8}),
\bfpage{211953}
(\byear{2022})
\doiurl{10.1098/rsos.211953}
\end{barticle}
\endbibitem

\bibitem[\protect\citeauthoryear{Ortiz-Sánchez
  et~al.}{2020}]{ortiz-sanchez_analysis_2020}
\begin{barticle}
\bauthor{\bsnm{Ortiz-Sánchez}, \binits{E.}},
\bauthor{\bsnm{Velando-Soriano}, \binits{A.}},
\bauthor{\bsnm{Pradas-Hernández}, \binits{L.}},
\bauthor{\bsnm{Vargas-Román}, \binits{K.}},
\bauthor{\bsnm{Gómez-Urquiza}, \binits{J.L.}},
\bauthor{\bsnm{Cañadas-De La~Fuente}, \binits{G.A.}},
\bauthor{\bsnm{Albendín-García}, \binits{L.}}:
\batitle{Analysis of the {Anti}-{Vaccine} {Movement} in {Social} {Networks}:
  {A} {Systematic} {Review}}.
\bjtitle{International Journal of Environmental Research and Public Health}
\bvolume{17}(\bissue{15}),
\bfpage{5394}
(\byear{2020})
\doiurl{10.3390/ijerph17155394}
\end{barticle}
\endbibitem

\bibitem[\protect\citeauthoryear{Roozenbeek
  et~al.}{2022}]{Roozenbeek2022Psychological}
\begin{barticle}
\bauthor{\bsnm{Roozenbeek}, \binits{J.}},
\bauthor{\bsnm{Linden}, \binits{S.}},
\bauthor{\bsnm{Goldberg}, \binits{B.}},
\bauthor{\bsnm{Rathje}, \binits{S.}},
\bauthor{\bsnm{Lewandowsky}, \binits{S.}}:
\batitle{Psychological inoculation improves resilience against misinformation
  on social media}.
\bjtitle{Science Advances}
\bvolume{8}(\bissue{34}),
\bfpage{6254}
(\byear{2022})
\doiurl{10.1126/sciadv.abo6254}
\end{barticle}
\endbibitem

\bibitem[\protect\citeauthoryear{Lu et~al.}{2023}]{Lu2024Psychological}
\begin{barticle}
\bauthor{\bsnm{Lu}, \binits{C.}},
\bauthor{\bsnm{Hu}, \binits{B.}},
\bauthor{\bsnm{Li}, \binits{Q.}},
\bauthor{\bsnm{Bi}, \binits{C.}},
\bauthor{\bsnm{Ju}, \binits{X.-D.}}:
\batitle{Psychological inoculation for credibility assessment, sharing
  intention, and discernment of misinformation: Systematic review and
  meta-analysis}.
\bjtitle{J Med Internet Res}
\bvolume{25},
\bfpage{49255}
(\byear{2023})
\doiurl{10.2196/49255}
\end{barticle}
\endbibitem

\bibitem[\protect\citeauthoryear{Kreps and Kriner}{2022}]{Kreps2022Infodemic}
\begin{barticle}
\bauthor{\bsnm{Kreps}, \binits{S.E.}},
\bauthor{\bsnm{Kriner}, \binits{D.L.}}:
\batitle{The covid-19 infodemic and the efficacy of interventions intended to
  reduce misinformation}.
\bjtitle{Public Opinion Quarterly}
\bvolume{86}(\bissue{1}),
\bfpage{162}--\blpage{175}
(\byear{2022})
\doiurl{10.1093/poq/nfab075}
{\href{https://arxiv.org/abs/https://academic.oup.com/poq/article-pdf/86/1/162/45442581/nfab075.pdf}{{https://academic.oup.com/poq/article-pdf/86/1/162/45442581/nfab075.pdf}}}
\end{barticle}
\endbibitem

\bibitem[\protect\citeauthoryear{BOWLES et~al.}{2025}]{Bowles2025Sustaining}
\begin{botherref}
\oauthor{\bsnm{BOWLES}, \binits{J.}},
\oauthor{\bsnm{CROKE}, \binits{K.}},
\oauthor{\bsnm{LARREGUY}, \binits{H.}},
\oauthor{\bsnm{LIU}, \binits{S.}},
\oauthor{\bsnm{MARSHALL}, \binits{J.}}:
Sustaining exposure to fact-checks: Misinformation discernment, media
  consumption, and its political implications.
American Political Science Review,
1--24
(2025)
\doiurl{10.1017/S0003055424001394}
\end{botherref}
\endbibitem

\bibitem[\protect\citeauthoryear{Butts et~al.}{2023}]{Butts2023Mathematical}
\begin{barticle}
\bauthor{\bsnm{Butts}, \binits{D.J.}},
\bauthor{\bsnm{Bollman}, \binits{S.A.}},
\bauthor{\bsnm{Murillo}, \binits{M.S.}}:
\batitle{Mathematical modeling of disinformation and effectiveness of
  mitigation policies}.
\bjtitle{Scientific Reports}
\bvolume{13}(\bissue{1}),
\bfpage{18735}
(\byear{2023})
\doiurl{10.1038/s41598-023-45710-2}
\end{barticle}
\endbibitem

\bibitem[\protect\citeauthoryear{Bak-Coleman
  et~al.}{2022}]{Bak-Coleman2022Combining}
\begin{barticle}
\bauthor{\bsnm{Bak-Coleman}, \binits{J.B.}},
\bauthor{\bsnm{Alfano}, \binits{M.}},
\bauthor{\bsnm{Barfuss}, \binits{W.}},
\bauthor{\bsnm{Bergstrom}, \binits{C.T.}},
\bauthor{\bsnm{Centeno}, \binits{M.A.}},
\bauthor{\bsnm{Couzin}, \binits{I.D.}},
\bauthor{\bsnm{Donges}, \binits{J.F.}},
\bauthor{\bsnm{Galesic}, \binits{M.}},
\bauthor{\bsnm{Gersick}, \binits{A.S.}},
\bauthor{\bsnm{Jacquet}, \binits{J.}}, \betal:
\batitle{Combining interventions to reduce the spread of viral misinformation}.
\bjtitle{Nature Human Behaviour}
\bvolume{6}(\bissue{11}),
\bfpage{1564}--\blpage{1574}
(\byear{2022})
\doiurl{10.1038/s41562-022-01453-1}
\end{barticle}
\endbibitem

\bibitem[\protect\citeauthoryear{DeVerna et~al.}{2024}]{DeVerna_2024}
\begin{barticle}
\bauthor{\bsnm{DeVerna}, \binits{M.R.}},
\bauthor{\bsnm{Aiyappa}, \binits{R.}},
\bauthor{\bsnm{Pacheco}, \binits{D.}},
\bauthor{\bsnm{Bryden}, \binits{J.}},
\bauthor{\bsnm{Menczer}, \binits{F.}}:
\batitle{Identifying and characterizing superspreaders of low-credibility
  content on twitter}.
\bjtitle{PLoS ONE}
\bvolume{19}(\bissue{5}),
\bfpage{0302201}
(\byear{2024})
\doiurl{10.1371/journal.pone.0302201}
\end{barticle}
\endbibitem

\bibitem[\protect\citeauthoryear{Brashier
  et~al.}{2021}]{Brashier2021TimingMatters}
\begin{barticle}
\bauthor{\bsnm{Brashier}, \binits{N.M.}},
\bauthor{\bsnm{Pennycook}, \binits{G.}},
\bauthor{\bsnm{Berinsky}, \binits{A.J.}},
\bauthor{\bsnm{Rand}, \binits{D.G.}}:
\batitle{Timing matters when correcting fake news}.
\bjtitle{Proceedings of the National Academy of Sciences}
\bvolume{118}(\bissue{5}),
\bfpage{2020043118}
(\byear{2021})
\doiurl{10.1073/pnas.2020043118}
{\href{https://arxiv.org/abs/https://www.pnas.org/doi/pdf/10.1073/pnas.2020043118}{{https://www.pnas.org/doi/pdf/10.1073/pnas.2020043118}}}
\end{barticle}
\endbibitem

\bibitem[\protect\citeauthoryear{DeVerna et~al.}{2021}]{covaxxy}
\begin{barticle}
\bauthor{\bsnm{DeVerna}, \binits{M.}},
\bauthor{\bsnm{Pierri}, \binits{F.}},
\bauthor{\bsnm{Truong}, \binits{B.}},
\bauthor{\bsnm{Bollenbacher}, \binits{J.}},
\bauthor{\bsnm{Axelrod}, \binits{D.}},
\bauthor{\bsnm{Loynes}, \binits{N.}},
\bauthor{\bsnm{Torres-Lugo}, \binits{C.}},
\bauthor{\bsnm{Yang}, \binits{K.-C.}},
\bauthor{\bsnm{Menczer}, \binits{F.}},
\bauthor{\bsnm{Bryden}, \binits{J.}}:
\batitle{{CoVaxxy: A Collection of English-Language Twitter Posts About
  COVID-19 Vaccines}}.
\bjtitle{Proc. Intl. AAAI Conf. on Web and Social Media (ICWSM)}
\bvolume{15}(\bissue{1}),
\bfpage{992}--\blpage{999}
(\byear{2021})
\doiurl{10.1609/icwsm.v15i1.18122}
\end{barticle}
\endbibitem

\bibitem[\protect\citeauthoryear{{Centers for Disease Control {and} Prevention
  (CDC)}}{2021}]{CDC2021}
\begin{botherref}
\oauthor{\bsnm{{Centers for Disease Control {and} Prevention (CDC)}}}:
{COVID-19 Vaccinations in the United States, County}.
Available online
(2021).
\url{https://data.cdc.gov/Vaccinations/COVID-19-Vaccinations-in-the-United-States-County/8xkx-amqh}
\end{botherref}
\endbibitem

\bibitem[\protect\citeauthoryear{M{\o}nsted and
  Lehmann}{2022}]{monstedAlgorithmicDetectionAnalysis2019}
\begin{barticle}
\bauthor{\bsnm{M{\o}nsted}, \binits{B.}},
\bauthor{\bsnm{Lehmann}, \binits{S.}}:
\batitle{{Characterizing polarization in online vaccine discourse --- A
  large-scale study}}.
\bjtitle{PLOS ONE}
\bvolume{17}(\bissue{2}),
\bfpage{0263746}
(\byear{2022})
\doiurl{10.1371/journal.pone.0263746}
\end{barticle}
\endbibitem

\bibitem[\protect\citeauthoryear{Vehtari
  et~al.}{2017}]{vehtariPracticalBayesianModel2017}
\begin{barticle}
\bauthor{\bsnm{Vehtari}, \binits{A.}},
\bauthor{\bsnm{Gelman}, \binits{A.}},
\bauthor{\bsnm{Gabry}, \binits{J.}}:
\batitle{Practical {{Bayesian}} model evaluation using leave-one-out
  cross-validation and {{WAIC}}}.
\bjtitle{Statistics and Computing}
\bvolume{27}(\bissue{5}),
\bfpage{1413}--\blpage{1432}
(\byear{2017})
\doiurl{10.1007/s11222-016-9696-4}
\end{barticle}
\endbibitem

\bibitem[\protect\citeauthoryear{Zhang et~al.}{2022}]{zhang-etal-2022-changes}
\begin{bchapter}
\bauthor{\bsnm{Zhang}, \binits{J.}},
\bauthor{\bsnm{DeLucia}, \binits{A.}},
\bauthor{\bsnm{Dredze}, \binits{M.}}:
\bctitle{Changes in tweet geolocation over time: A study with carmen 2.0}.
In: \bbtitle{Proceedings of the Eighth Workshop on Noisy User-generated Text
  (W-NUT 2022)},
pp. \bfpage{1}--\blpage{14}.
\bpublisher{Association for Computational Linguistics},
\blocation{Gyeongju, Republic of Korea}
(\byear{2022}).
\burl{https://aclanthology.org/2022.wnut-1.1/}
\end{bchapter}
\endbibitem

\bibitem[\protect\citeauthoryear{Liu et~al.}{2019}]{liu2019roberta}
\begin{barticle}
\bauthor{\bsnm{Liu}, \binits{Y.}},
\bauthor{\bsnm{Ott}, \binits{M.}},
\bauthor{\bsnm{Goyal}, \binits{N.}},
\bauthor{\bsnm{Du}, \binits{J.}},
\bauthor{\bsnm{Joshi}, \binits{M.}},
\bauthor{\bsnm{Chen}, \binits{D.}},
\bauthor{\bsnm{Levy}, \binits{O.}},
\bauthor{\bsnm{Lewis}, \binits{M.}},
\bauthor{\bsnm{Zettlemoyer}, \binits{L.}},
\bauthor{\bsnm{Stoyanov}, \binits{V.}}:
\batitle{{RoBERTa: A Robustly Optimized BERT Pretraining Approach}}.
\bjtitle{Preprint arXiv:1907.11692}
(\byear{2019})
\doiurl{10.48550/arXiv.1907.11692}
\end{barticle}
\endbibitem

\bibitem[\protect\citeauthoryear{Chicco
  et~al.}{2021}]{chiccoMatthewsCorrelationCoefficient2021}
\begin{barticle}
\bauthor{\bsnm{Chicco}, \binits{D.}},
\bauthor{\bsnm{T{\"o}tsch}, \binits{N.}},
\bauthor{\bsnm{Jurman}, \binits{G.}}:
\batitle{The {{Matthews}} correlation coefficient ({{MCC}}) is more reliable
  than balanced accuracy, bookmaker informedness, and markedness in two-class
  confusion matrix evaluation}.
\bjtitle{BioData Mining}
\bvolume{14}(\bissue{1}),
\bfpage{13}
(\byear{2021})
\doiurl{10.1186/s13040-021-00244-z}
\end{barticle}
\endbibitem

\bibitem[\protect\citeauthoryear{Efron and
  Tibshirani}{1993}]{efronIntroductionBootstrap1993}
\begin{bbook}
\bauthor{\bsnm{Efron}, \binits{B.}},
\bauthor{\bsnm{Tibshirani}, \binits{R.J.}}:
\bbtitle{An {{Introduction}} to the {{Bootstrap}}}.
\bpublisher{{Springer US}},
\blocation{{Boston, MA}}
(\byear{1993}).
\doiurl{10.1007/978-1-4899-4541-9}
\end{bbook}
\endbibitem

\bibitem[\protect\citeauthoryear{Hamra et~al.}{2013}]{hamra_markov_2013}
\begin{barticle}
\bauthor{\bsnm{Hamra}, \binits{G.}},
\bauthor{\bsnm{MacLehose}, \binits{R.}},
\bauthor{\bsnm{Richardson}, \binits{D.}}:
\batitle{{Markov Chain Monte Carlo: An introduction for epidemiologists}}.
\bjtitle{International Journal of Epidemiology}
\bvolume{42}(\bissue{2}),
\bfpage{627}--\blpage{634}
(\byear{2013})
\doiurl{10.1093/ije/dyt043}
\end{barticle}
\endbibitem

\bibitem[\protect\citeauthoryear{Hoffman and
  Gelman}{2014}]{hoffmanNoUTurnSamplerAdaptively}
\begin{barticle}
\bauthor{\bsnm{Hoffman}, \binits{M.D.}},
\bauthor{\bsnm{Gelman}, \binits{A.}}:
\batitle{{The No-U-Turn Sampler: Adaptively Setting Path Lengths in Hamiltonian
  Monte Carlo}}.
\bjtitle{Journal of Machine Learning Research}
\bvolume{15}(\bissue{47}),
\bfpage{1593}--\blpage{1623}
(\byear{2014})
\end{barticle}
\endbibitem

\bibitem[\protect\citeauthoryear{Phan et~al.}{2019}]{phan2019composable}
\begin{barticle}
\bauthor{\bsnm{Phan}, \binits{D.}},
\bauthor{\bsnm{Pradhan}, \binits{N.}},
\bauthor{\bsnm{Jankowiak}, \binits{M.}}:
\batitle{{Composable Effects for Flexible and Accelerated Probabilistic
  Programming in NumPyro}}.
\bjtitle{Preprint arXiv:1912.11554}
(\byear{2019})
\doiurl{10.48550/arXiv.1912.11554}
\end{barticle}
\endbibitem

\bibitem[\protect\citeauthoryear{{Centers for Disease Control and
  Prevention}}{2023}]{cdc_isolation_duration}
\begin{botherref}
\oauthor{\bsnm{{Centers for Disease Control and Prevention}}}:
{Ending Isolation and Precautions for People with COVID-19: Interim Guidance}
(2023).
\url{https://www.cdc.gov/coronavirus/2019-ncov/hcp/duration-isolation.html}
\end{botherref}
\endbibitem

\bibitem[\protect\citeauthoryear{ASPE}{2021}]{aspe_vaccine_hesitancy}
\begin{botherref}
\oauthor{\bsnm{ASPE}}:
{Vaccine Hesitancy for COVID-19: State, County, and Local Estimates}
(2021).
\url{https://aspe.hhs.gov/reports/vaccine-hesitancy-covid-19-state-county-local-estimates}
\end{botherref}
\endbibitem

\bibitem[\protect\citeauthoryear{Dehning et~al.}{2020}]{dehning_inferring_2020}
\begin{barticle}
\bauthor{\bsnm{Dehning}, \binits{J.}},
\bauthor{\bsnm{Zierenberg}, \binits{J.}},
\bauthor{\bsnm{Spitzner}, \binits{F.P.}},
\bauthor{\bsnm{Wibral}, \binits{M.}},
\bauthor{\bsnm{Neto}, \binits{J.P.}},
\bauthor{\bsnm{Wilczek}, \binits{M.}},
\bauthor{\bsnm{Priesemann}, \binits{V.}}:
\batitle{{Inferring change points in the spread of COVID-19 reveals the
  effectiveness of interventions}}.
\bjtitle{Science}
\bvolume{369}(\bissue{6500}),
\bfpage{9789}
(\byear{2020})
\doiurl{10.1126/science.abb9789}
\end{barticle}
\endbibitem

\bibitem[\protect\citeauthoryear{Marinov and
  Marinova}{2022}]{marinov_adaptive_2022}
\begin{barticle}
\bauthor{\bsnm{Marinov}, \binits{T.T.}},
\bauthor{\bsnm{Marinova}, \binits{R.S.}}:
\batitle{Adaptive {SIR} model with vaccination: simultaneous identification of
  rates and functions illustrated with {COVID}-19}.
\bjtitle{Scientific Reports}
\bvolume{12}(\bissue{1}),
\bfpage{15688}
(\byear{2022})
\doiurl{10.1038/s41598-022-20276-7}
\end{barticle}
\endbibitem

\bibitem[\protect\citeauthoryear{Tenforde
  et~al.}{2022}]{cdc_covid_vaccine_effectiveness}
\begin{barticle}
\bauthor{\bsnm{Tenforde}, \binits{M.W.}},
\bauthor{\bsnm{Self}, \binits{W.H.}},
\bauthor{\bsnm{Gaglani}, \binits{M.}},
\bauthor{\bsnm{al.}}:
\batitle{{Effectiveness of mRNA Vaccination in Preventing COVID-19–Associated
  Invasive Mechanical Ventilation and Death — United States, March
  2021–January 2022}}.
\bjtitle{Morbidity and Mortality Weekly Report}
\bvolume{71}(\bissue{12}),
\bfpage{459}--\blpage{465}
(\byear{2022})
\doiurl{10.15585/mmwr.mm7112e1}
\end{barticle}
\endbibitem

\bibitem[\protect\citeauthoryear{Lopez~Bernal
  et~al.}{2021}]{Bernal2021_covid_efficacy}
\begin{barticle}
\bauthor{\bsnm{Lopez~Bernal}, \binits{J.}},
\bauthor{\bsnm{Andrews}, \binits{N.}},
\bauthor{\bsnm{Gower}, \binits{C.}},
\bauthor{\bsnm{Gallagher}, \binits{E.}},
\bauthor{\bsnm{Simmons}, \binits{R.}},
\bauthor{\bsnm{Thelwall}, \binits{S.}},
\bauthor{\bsnm{Stowe}, \binits{J.}},
\bauthor{\bsnm{Tessier}, \binits{E.}},
\bauthor{\bsnm{Groves}, \binits{N.}},
\bauthor{\bsnm{Dabrera}, \binits{G.}},
\bauthor{\bsnm{Myers}, \binits{R.}},
\bauthor{\bsnm{Campbell}, \binits{C.N.J.}},
\bauthor{\bsnm{Amirthalingam}, \binits{G.}},
\bauthor{\bsnm{Edmunds}, \binits{M.}},
\bauthor{\bsnm{Zambon}, \binits{M.}},
\bauthor{\bsnm{Brown}, \binits{K.E.}},
\bauthor{\bsnm{Hopkins}, \binits{S.}},
\bauthor{\bsnm{Chand}, \binits{M.}},
\bauthor{\bsnm{Ramsay}, \binits{M.}}:
\batitle{Effectiveness of covid-19 vaccines against the b.1.617.2 (delta)
  variant}.
\bjtitle{New England Journal of Medicine}
\bvolume{385}(\bissue{7}),
\bfpage{585}--\blpage{594}
(\byear{2021})
\doiurl{10.1056/NEJMoa2108891} .
\bcomment{PMID: 34289274}
\end{barticle}
\endbibitem

\end{thebibliography}

\newpage
\begin{appendices}

\section{Supplementary Methods}

\renewcommand{\thefigure}{S\arabic{figure}}
\renewcommand{\thetable}{S\arabic{table}}

\subsection{Data}

We use three main data sources in this work: the CoVaxxy Tweets database, which contains Tweets related to COVID vaccines; CDC records of COVID cases and vaccinations in US counties; and Mønsted and Lehmann's antivaccine tweets dataset, which supplements our own labeled data for training a tweet classifier. Details are provided below.

The primary data source for this work is the CoVaxxy project~\cite{covaxxy} from Indiana University's Observatory on Social Media.  CoVaxxy collects tweets related to COVID vaccines and vaccine hesitancy, and geolocates the tweets to US counties when possible. The tweets were geolocated using Carmen~\cite{zhang-etal-2022-changes}, a tool which looks for geolocation meta data in tweets and self-identified locations in user profiles and post texts. Social bots were deliberately not excluded from our data, as bots may also contribute to population exposures to antivaccine content. Using this dataset, we can track the online discourse surrounding COVID vaccination in individual US counties. 

The US Center for Disease Control (CDC) collected and published data on COVID health outcomes and vaccination in US counties, including cumulative cases, deaths, and vaccinations~\cite{CDC2021} for each county, for each day. This is the source of the public health metrics in this work. Due to differences in local reporting systems and reporting schedules, some entries in the cumulative counts are imputed by CDC as the last known value until they are updated with new reports from the counties or states. 
Counties within the state of Texas are excluded from our dataset because the official CDC data on COVID vaccinations does not contain information about Texas counties until after October 22, 2021.

To train our antivaccine tweet classifier, we require a dataset of tweets labeled as ``antivaccine'' or ``other.'' We supplement our own labeled data with a similar dataset created by Mønsted and Lehmann \cite{monstedAlgorithmicDetectionAnalysis2019}. Restrictions apply to the availability of this dataset, which therefore is not publicly available. The data is, however, available from the authors upon reasonable request. Although this dataset is not specific to COVID antivaccine sentiment, we found its content to be similar enough to improve the performance of our classifier. We used the Mønsted and Lehmann dataset for training our model but not as part of test data used to evaluate our classifier's performance.

\subsection{Antivaccine Tweet Classifier} 

To track the volume of antivaccine content produced in each county, we built a text classifier that determines if a tweet is expressing antivaccine sentiment. The classifier takes the text of a tweet as input and returns a label, either ``antivaccine'' or ``other.''  The classifier is a pre-trained neural network based on the RoBERTa language model~\cite{liu2019roberta}, and is fine-tuned for our classification problem on a set of manually labeled tweets.

Our labeled training and test data were coded by two human annotators and examined for agreement. Cases where annotators disagreed were discarded. To be labeled as antivaccine, a tweet must ``express the belief that safe, effective COVID vaccines are bad, ineffective, not actually a vaccine, or harmful (without specific evidence).'' Most tweets expressing the belief that COVID vaccines are harmful made one of a few common claims, so these common claims were manually checked using reputable fact-checkers (e.g., PolitiFact, FactCheck.org) and CDC publications; in general, the common claims of harm were found to be false and labeled as antivaccine. We did not code other possible categories of vaccine sentiment, such as ``pro-vaccination,'' because our goal was to analyze the effect of antivaccine tweets.

The classifier was trained on 4,200 labeled tweets from the CoVaxxy dataset. To increase training data volume and variety, we also added to this training set 2,000 tweets randomly selected from the Mønsted and Lehmann dataset~\cite{monstedAlgorithmicDetectionAnalysis2019}, which were labeled by three human annotators; cases where annotators disagreed were discarded. The model was tested on a hold-out set of 900 labeled tweets from the CoVaxxy dataset, labeled by the same method and annotators as the training data. We did not use cross-validation because (i) the training data includes Mønsted and Lehmann's tweets, which do not match 2021 COVID-related tweets, and (ii) our sample of training tweets from the CoVaxxy dataset deliberately included a greater proportion of antivaccine tweets to help with classifier learning. Instead we produced the test dataset by pure random sampling from the CoVaxxxy data to ensure that classification metrics like $F_1$ are unbiased.

The classifier was evaluated using common classification metrics. It has an accuracy of $0.94 \pm 0.02$, an $F_1$ of $0.74 \pm 0.07$, and a Matthews Correlation Coefficient~\cite{chiccoMatthewsCorrelationCoefficient2021} of $0.72 \pm 0.08$, where 95\% confidence intervals are computed by the bootstrap procedure~\cite{efronIntroductionBootstrap1993}. 
This classifier was used to determine the number of antivaccine tweets geolocated in each county on each date. While some individual tweets may be misclassified, we believe the performance of the classifier is adequate to determine population-level trends and relative magnitudes in the prevalence of antivaccine tweets.

\subsection{Estimating Parameters}

Our goal is to infer the parameters of the SIRVA model from the data.  In particular, we want to infer $\gamma_e$ to understand whether exposure to antivaccine tweets leads additional people to become unwilling to be vaccinated. To estimate the likely range of the model parameters, we use Bayesian inference with MCMC sampling~\cite{hamra_markov_2013}. The goal of Bayesian inference is to find the probability distribution (which we call the posterior distribution) that describes the likely values of a parameter given the data. From the posterior distributions, we can get a mean estimate, the 95\% high probability interval (``HDI''), and a $p$-value for each parameter. 

At a high level, Bayesian MCMC inference samples random parameter values proportionally to their likelihood given the data. The MCMC algorithm begins by sampling from a prior distribution, which is defined by specifying plausible ranges of the parameters, and gradually converges to the posterior distribution. We use the standard NUTS algorithm~\cite{hoffmanNoUTurnSamplerAdaptively} implemented in the NumPyro package~\cite{phan2019composable}. 

To specify a likelihood function for the SIRVA model given the data, we will define probability distributions for the daily changes in each of our observed variables: cumulative cases ($C=I+R$), vaccinations ($V$), susceptible individuals ($S=N-(V+C)$), and recovered individuals ($R$). We assume $R_t \approx C_{t-8}$, based on a typical time from initial symptoms to non-infectious status of 10 days~\cite{cdc_isolation_duration} and a lag time between initial symptoms and a positive test of about 2 days.

We want to define the probability of observing the daily changes in cases, vaccinations, recoveries, and susceptibles (i.e., $x_j = \Delta C, \Delta V, \Delta R, -\Delta S$) given estimates of these data ($\mu_j$) from the SIRVA model equations using the sampled parameters. We define the probability of observing data $x_j$ given estimate $\mu_j$ with the negative binomial distribution, which accounts for noise in the data through a concentration parameter $\phi_j$:
\begin{align}\label{eq:negbin_dist}
P_{\text{NegBin}}(x_j \mid \mu_j, \phi_j)  = \binom{x_j + \phi_j - 1}{x_j} \,
\left( \frac{\mu_j}{\mu_j+\phi_j} \right)^{x_j} \, \left(
\frac{\phi_j}{\mu_j+\phi_j} \right)^{\phi_j}.
\end{align}
Each observed variable is distributed according to the negative binomial, for example, $
    \Delta C \sim P_{\text{NegBin}}(x_C \mid \mu_C, \phi_C) 
$.
The $\phi$ parameters are inferred by the MCMC sampler in the same way as other model parameters.

We define the likelihood function of our parameters given the data as the product of these four negative binomial distributions: 
\begin{align} \label{eq:likelihood}
    L = \sum_{j} \sum_{i} \log( \, P_{\text{NegBin}}(x_{ij} \mid \mu_j, \phi_j) \,)
\end{align}
where the index $i$ represents a data point for a particular date and county.

The MCMC sampler also requires us to specify prior distributions for each model parameter we want to infer. For the basic epidemic parameters, we use normal distributions as priors:
\begin{align*}
    \rho &\sim \text{Norm}(\mu=0.1, \sigma=0.3) \\
    \beta &\sim \text{Norm}(\mu=0.2, \sigma=0.3) \\
    \nu &\sim \text{Norm}(\mu=0.0025, \sigma=0.1).
\end{align*}
These distributions have high variance relative to the plausible parameter ranges so that the priors do not strongly influence the final inferred parameter values. We note that $\nu$ has a somewhat sharper prior than other parameters, informed by knowing the approximate number of total vaccinations in the studied time period. This sharper prior was necessary to eliminate identifiability issues between $\nu$ and $\gamma_e$.

For the vaccine-hesitancy parameters, $\gamma_e$ and $\gamma_p$, we choose priors centered at zero. This way, if the posterior is found to be non-zero, we know the prior did not bias that conclusion. For $\alpha_0$, we choose a weak prior based on general estimates of vaccine hesitancy in the population~\cite{aspe_vaccine_hesitancy}. We use these priors:
\begin{align*}
    \gamma_e &\sim \text{Norm}(\mu=0, \sigma=1) \\
    \gamma_p &\sim \text{Norm}(\mu=0, \sigma=0.5) \\
    \alpha_0 &\sim \text{Norm}(\mu=0.2, \sigma=0.5).
\end{align*}

Finally, the concentration parameters defined above are given weak Gamma priors, which flexibly capture variance in the data:
\begin{align*}
    \phi_j \sim \text{Gamma}(\phi_j \mid a, b) = \frac{b^{a}}{\Gamma(a)} \, \phi_j^{a - 1}
\end{align*}
with parameters $a=1$ and $b=6$.

We set weak upper and lower bounds for all the priors to prevent runtime errors associated with negative or very large parameter values. In particular, the parameters that must be positive ($\rho$, $\beta$, $\nu$, $A_0$) are restricted to be positive definite with a lower bound of $10^{-15}$. Upper bounds are set well above realistic ranges, e.g., $1.0$ for $A_0$, the initial fraction of the population unwilling to be vaccinated.

In our model, the parameters $\nu$ and $\beta$ can vary over time, following prior work~\cite{dehning_inferring_2020, marinov_adaptive_2022}. The $\nu$ parameter is allowed to change once every six weeks to account for changing nationwide vaccine availability. The $\beta$ parameter is allowed to change once every three weeks to account for changing mean reproduction numbers associated with the emergence of new variants (e.g., the Delta variant late in our observation window) and changing public health policies (e.g., the imposition or lifting of lockdowns and mask mandates). The multiple values of these parameters are inferred by the MCMC sampler just like the other parameters.

Our dataset comprises 1,319 counties and 188 dates for which we have sufficient Twitter and public health data. We use a subsample of this data to perform our inferences. We sample by date to minimize the effect of temporal autocorrelations in the data; specifically, we use every 8th day in the time series data for each county. This interval is chosen for two reasons: (i)~it is slightly more than a week, so it smooths over data lags associated with weekly reporting cycles, and (ii)~it is the approximate recovery time used to compute $R$. We also randomly sample by county to limit the number of model parameters (each county $i$ introduces an additional parameter $\alpha_{0,i}$). Our Bayesian MCMC inference tools struggle with numerical stability when the numbers of parameters and data points get too large. We therefore use 400 randomly selected counties and 24 evenly-spaced dates from our data. Although this sampling ultimately reduces the precision of our inferred parameter values, we do not believe it introduces any systematic biases; results are consistent across different random samples of the counties.

\subsection{Constructing the Causal Graphical Model}

\begin{figure}
\centering
  \includegraphics[width=0.8\textwidth]{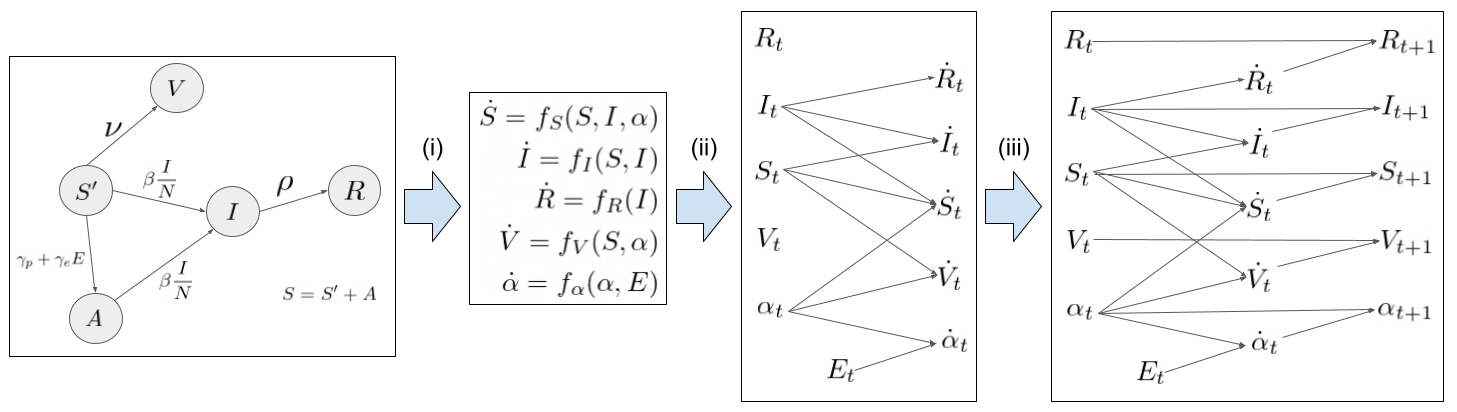}
\caption{Steps to construct the causal graphical model from the SIRVA compartmental model.}
\label{fig:CGM_construction}
\end{figure}

\begin{figure}
\centering
  \includegraphics[width=0.8\textwidth]{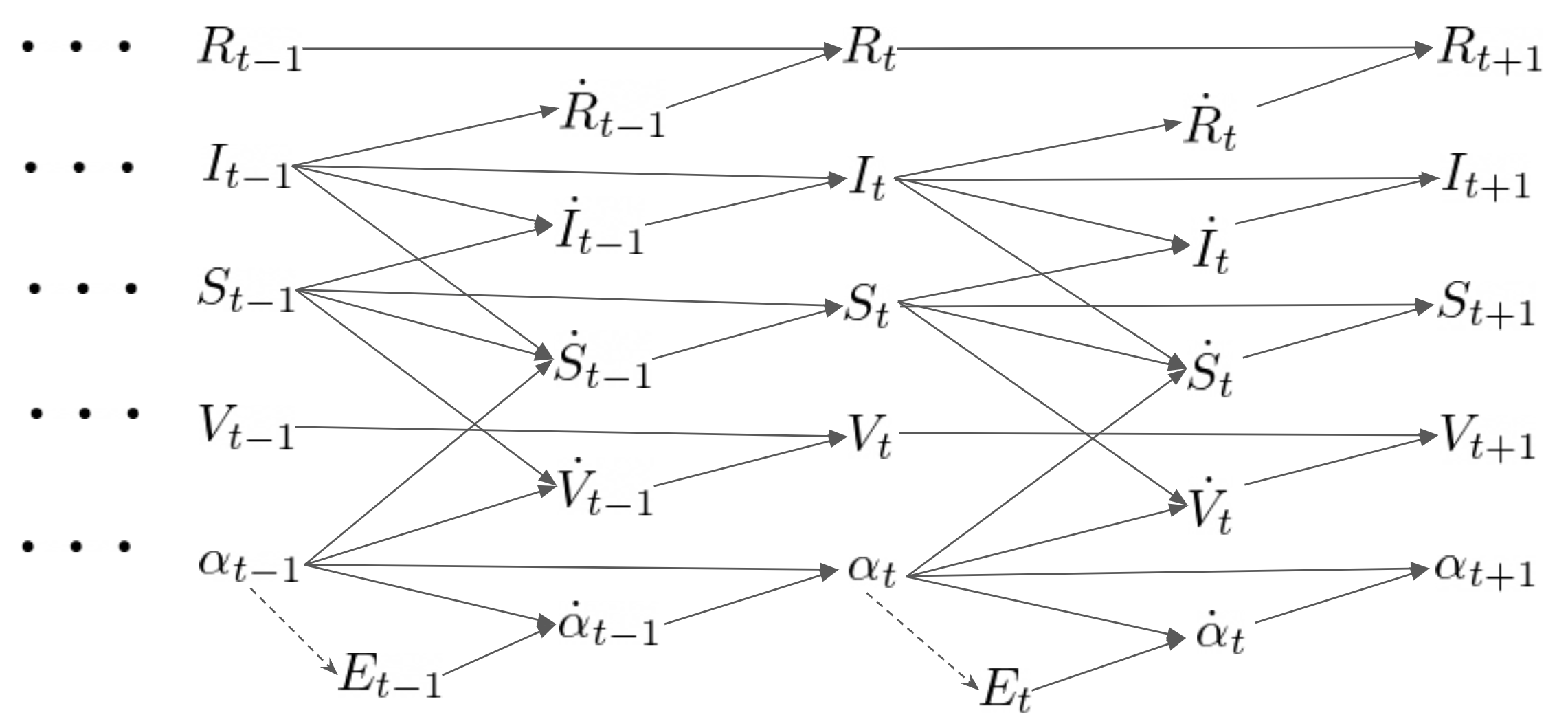}
\caption{Causal graphical model (CGM) corresponding to the SIRVA model.  The diagram shows the causal antecedents of the model variables at time $t$. Here the notation $\dot{X}$ denotes the quantity $\frac{dX}{dt}$, and each arrow represents a causal relationship, where the source variable (on the left) causes the target variable (on the right). The dotted arrows denote an additional possible confounding relationship from $\alpha$ to $E$ in the model; this relationship is assumed to exist in the CGM, and our Average Treatment Effect accounts for its presence. Note that the causal graph is a lattice that can be extended backward in time to an initial state (e.g., $\alpha_0$). 
}
\label{fig:causal_graphical_model}
\end{figure}

We construct a Causal Graphical Model (CGM) corresponding to the SIRVA model in a few steps, illustrated in Supplementary Figure~\ref{fig:CGM_construction}. (i) We consider the derivatives of the key dynamic variables ($\dot{S}$, $\dot{I}$, $\dot{R}$, $\dot{V}$, $\dot{\alpha}$), and identify the variables on which they functionally depend in the model equations. (ii) We construct a simple bipartite graph from the variables to their derivatives, where each arrow represents a dependency. (iii) We complete a full time step in the CGM by creating additional links from the variables and their derivatives at time $t$ to the variables at time $t+1$. 

Supplementary Figure~\ref{fig:causal_graphical_model} illustrates the final CGM, which includes additional dotted lines representing a confounding relationship from $\alpha$ to $E$ not explicitly captured by the SIRVA model equations. This potential confounding relationship is included because vaccine-hesitant people may be more likely to engage with antivaccine content online, increasing their exposure. The chain of time steps is also extended backward in time to include the variables at times $t-1, \dots, 0$, back to the initial conditions (e.g., $\alpha_0$).

\subsection{Deriving the ATE Estimand}

We leverage the causal graphical model constructed in the previous supplementary methods section to find the average treatment effect (ATE) of exposure per capita $E$ on the vaccine uptake rate per capita $\dot{V_i}/N_i$. 
The CGM of Supplementary Figure~\ref{fig:causal_graphical_model} shows a path $E_{t-1} \rightarrow \dot{\alpha}_{t-1} \rightarrow \alpha_{t} \rightarrow \dot{V}_t$, indicating that there is a causal chain of effects from $E$ to $\dot{V}$. In addition, there is a causal chain from $\alpha_{t-1}$ to both $\dot{V}$ and $E$, indicating that $\alpha_{t-1}$ is a confounding variable that must be considered. To account for these relationships, we use do-calculus~\cite{pearlDoCalculusRevisited2012}, as implemented in the DoWhy python package, to find the generic form of the ATE:
\begin{align*}
\frac{d}{dE_{t-1}} \Big(\mathbb{E}\Big[\frac{\dot{V}_{t,i}}{N_i} \mid \alpha_{t-1}\Big]\Big)
\end{align*}
where $\mathbb{E}[x]$ denotes the expectation value of $x$ over the data and the posterior samples.

To write this expression in terms of our model variables and parameters, we plug in the expression for $\dot{V_t}$ in terms of $E_{t-1}$ according to our model equations. We can look at the causal path through the CGM ($E_{t-1} \rightarrow \dot{\alpha}_{t-1} \rightarrow \alpha_{t-1} \rightarrow \dot{V}_t$) to find the relevant model equations (Eqs.~\ref{eq:SIRVA_model_Vdot}, \ref{eq:SIRVA_model_alphadot}, \ref{eq:SIRVA_model_alpha}):
\begin{align*}
    \dot{V}_{t}     = \nu S_{t-1} (1-\alpha_{t-1})\\
    \dot{\alpha}_{t-1} = (\gamma_p + \gamma_e E_{t-1})(1-\alpha_{t-1}) \\
    \alpha_{t}      \approx  \dot{\alpha}_{t-1}\Delta t + \alpha_{t-1}.
\end{align*}
Plugging in these expressions and $\Delta t = 1$ for a one time step change, we get:
\begin{align*}
    \dot{V}_{t}  \approx \nu S_{t-1} (1-((\gamma_p + \gamma_e E_{t-1})(1-\alpha_{t-1}) + \alpha_{t-1})).
\end{align*}
Next, we need to take the expectation value over our data set and the joint posterior distribution of our parameters. Specifically, we compute an expectation value over our counties $i$, times $t$, and posterior samples $s$, and weight the expectation by county population, $N_i$. So the expression for $\mathbb{E}_{t,i,s}[\frac{\dot{V}_{t,i}}{N_i}]$ is rewritten as:
\begin{multline*}
    \mathbb{E}_{t,i, s}\Big[\frac{\dot{V}_{t,i}}{N_i}\Big] \approx \frac{1}{(n_t-1) n_s (\sum_i N_i)} \\ \sum_{t=1}^{n_t} \sum_i \sum_s N_i \frac{(\nu_s S_{t-1,i} (1-((\gamma_{p,s} + \gamma_{e,s} E_{t-1, i})(1-\alpha_{t-1, i,s}) + \alpha_{t-1, i,s})))}{N_i}
\end{multline*}
where $n_t$ and $n_s$ are the number of dates and posterior samples, respectively. Note that this quantity depends explicitly on $\alpha_{t-1}$, so, $\mathbb{E}[\frac{\dot{V}_{t,i}}{N_i}\mid \alpha_{t-1}]  = \mathbb{E}_{t,i, s}[\frac{\dot{V}_{t,i}}{N_i}]$. 
Note also that the factor of $\frac{N_i}{N_i}=1$ results from the population-weighted averaging and our use of per-capita vaccination rates.

Taking the derivative with respect to $E_{t-1}$ and collapsing the expectation value to a more concise notation, we get our estimand:
\begin{align}
    \frac{d}{dE_{t-1}} \Big(\mathbb{E}\Big[\frac{\dot{V}_{t,i}}{N_i} \mid \alpha_{t-1}\Big]\Big) 
  &\approx - \frac{1}{(n_t-1) n_s (\sum_i N_i)} \sum_{t=1}^{n_t} \sum_i \sum_s \nu_s S_{t-1,i} \gamma_{e,s} (1-\alpha_{t-1, i,s}) \nonumber\\
 &= - \mathbb{E}_{t,i, s} \Big[\frac{1}{N_i}\nu_s S_{t-1,i} \gamma_{e,s} (1-\alpha_{t-1, i, s}) \Big] \nonumber\\
 &= - \mathbb{E} \Big[\frac{1}{N_i}\nu S_{t-1} \gamma_{e} (1-\alpha_{t-1}) \Big]. \label{eq:dVdotdE}
\end{align}
Note that in the second line of the above equation, we divide by $N_i$ because our expectation operation $\mathbb{E}$ is weighted by $N_i$.

We can estimate the change in vaccine uptake in each county on each date using a linear approximation of the total effect. This approximation is valid under the assumption that the total change in vaccination is small. (Below we justify this assumption by estimating the change with another, independent method.) Using this approximation, we can write the total change in vaccination per day given exposure as
\begin{align}
   \frac{\Delta  \dot{V}_{t,i}}{N_i} &\approx E_{t-1,i} \Big(\frac{d}{dE_{t-1}} \Big(\mathbb{E}_{t,i}\Big[\frac{\dot{V}_{t,i}}{N_i}\Big]\Big) \Big) \label{eq:DeltaVdotpercapita}
\end{align}
and the total change in vaccine uptake over the whole time period ($\Delta t$) and nation-wide population ($N$) as
\begin{align}
   {\Delta V} = N \Delta t \Big(\mathbb{E}_{t,i}\Big[\frac{{\Delta \dot{V}}_{t,i}}{N_i}\Big]\Big). \label{eq:DeltaV}
\end{align} 
Now plugging in Eqs.~\ref{eq:dVdotdE} and \ref{eq:DeltaVdotpercapita} into Eq.~\ref{eq:DeltaV}, we get our final result:
\begin{align}
\Delta V 
&= N \Delta t ~ \mathbb{E}_{t,i}\Big[E_{t-1,i} \Big(\frac{d}{dE_{t-1}} \Big(\mathbb{E}_{t,i}\Big[\frac{\dot{V}_{t,i}}{N_i} \mid \alpha_{t-1}\Big]\Big) \Big)\Big] \nonumber\\
&= - N \Delta t ~ \mathbb{E}_{t,i} \Big[ E_{t-1,i} \Big( \mathbb{E}_{t,i} \Big[\frac{1}{N_i}\nu S_{t-1} \gamma_{e} (1-\alpha_{t-1}) \Big]\Big)\Big]
\end{align}
where $N$ is the total population (about 337 million people in the US), $\Delta t$ is the length of the observation period in days (192 days from February to August 2021), and all expectation operations $\mathbb{E}$ are weighted by populations $N_i$.
This results in a figure of 14,086 vaccinations prevented between February and August 2021 in the US.

We also computed this quantity in an alternative way by simulating the SIRVA model with the inferred parameters in a counterfactual scenario where the exposure in all counties was set to zero. 
Comparing the real data with this scenario allowed us to infer an alternative estimate compatible with that obtained with the ATE-based method, helping validate our assumption about the small effect size. 
However, the ATE-based method is better able to explicitly account for potential confounding factors. 

\subsection{Estimating Case and Death Rates among the Unvaccinated Population}

We want to know the probability that a person who was unvaccinated would have become infected with COVID during the observation period from February to August 2021 in the United States. We can estimate this probability from the case and vaccination data, with the help of an estimate of vaccine effectiveness. Consider the probability that any person would have been infected, $P(C)$. We can break this probability down into two parts: the cases of vaccinated people $P(C,V)$ and the cases of unvaccinated people $P(C,\bar{V})$:
\begin{align*}
    P(C) = P(C,V) + P(C,\bar{V}) = P(C \mid V)P(V) + P(C \mid \bar{V})P(\bar{V}).
\end{align*}
We can relate these two parts using the effectiveness of vaccinations at preventing cases, defined as $\lambda_C = 1- \frac{P(C \mid V)}{P(C \mid \bar{V})}$:
\begin{align*}
    P(C) = (P(C\mid\bar{V})(1-\lambda_C)) P(V)+P(C\mid\bar{V})P(\bar{V}).
\end{align*}
We can solve for $P(C\mid\bar{V})$, the probability that an unvaccinated person would become infected: 
\begin{align*}
    P(C\mid\bar{V}) = P(C)[P(V)(1-\lambda_C) + P(\bar{V})]^{-1}. 
\end{align*}
We can get a real value of this quantity by plugging in mean estimates of these probabilities for the cases that occurred over a particular time period (e.g., the previous day) and the number of vaccinations at that point in time:
\begin{align*}
    P_t(C\mid\bar{V}) = \frac{\Delta C_t}{N-C}\Big[\frac{V_t}{N-C}(1-\lambda_C) + \frac{N-C-V_t}{N-C}\Big]^{-1}. 
\end{align*}
Summing over time, we find the total risk is:
\begin{align*}
    P(C\mid\bar{V}) = \sum_{t} \frac{\Delta C_t}{N-C}\Big[\frac{V_t}{N-C}(1-\lambda_C) + \frac{N-C-V_t}{N-C}\Big]^{-1}. 
\end{align*}
We can similarly derive the risk of an unvaccinated person dying using the effectiveness of vaccines at preventing deaths, $\lambda_D$:
\begin{align*}
    P(D\mid\bar{V}) = \sum_{t} \frac{\Delta D_t}{N-D}\Big[\frac{V_t}{N-D}(1-\lambda_D) + \frac{N-D-V_t}{N-D}\Big]^{-1}. 
\end{align*}
Computing these risks for an unvaccinated person in the United States over the period February-August 2021 using CDC data for cases, deaths, and vaccinations and values of $\lambda_C=0.93$ and $\lambda_D=0.94$~\cite{cdc_covid_vaccine_effectiveness, Bernal2021_covid_efficacy}, we find $P(C\mid\bar{V}) = 0.0387$ and $P(D\mid\bar{V}) = 0.00057$. This equates to 3,870 cases and 57 deaths per 100,000 unvaccinated people, which we use in the main text to estimate the numbers of cases and deaths resulting from exposure to antivaccine content on Twitter.

\end{appendices}

\end{document}